\documentclass[letter,longauth]{aa}  
\usepackage[varg]{txfonts}
\usepackage{graphicx}
\usepackage{lscape}
\usepackage{amsmath}
\usepackage{amssymb}
\usepackage{color}
\usepackage{natbib}
\bibpunct{(}{)}{;}{a}{}{,}
\usepackage{url}
\usepackage{multirow}
\usepackage{threeparttable}
\usepackage{booktabs} 
\usepackage{soul}
\usepackage{float}
\usepackage[colorlinks=true]{hyperref}
\usepackage{tablefootnote}
\usepackage{subcaption}
\hypersetup{colorlinks=true,allcolors=blue,citecolor=blue}
\usepackage{tabularx}
\usepackage[font=small,labelfont=bf]{caption}
\usepackage{orcidlink}
\usepackage{siunitx}
\usepackage{lastpage}

\titlerunning{Uncovering temperate worlds: an updated look at the K2-3 system with ESPRESSO}
\authorrunning{Srivastava et al.}

\begin{document}

   \title{Uncovering temperate worlds: an updated look at the K2-3 system with ESPRESSO}

\author{
Avidaan Srivastava\inst{1,*}\orcidlink{0009-0009-7136-1528},
Emma Pimentel\inst{1},
Am\'elie Gressier\inst{1}\orcidlink{0000-0003-0854-3002},
Ren\'e Doyon\inst{1,2}\orcidlink{0000-0001-5485-4675},
Melissa J. Hobson\inst{3,4}\orcidlink{0000-0002-5945-7975},
Alessandro Sozzetti\inst{5}\orcidlink{0000-0002-7504-365X},
Mario Damasso\inst{5}\orcidlink{0000-0001-9984-4278},
Yann Alibert\inst{6}\orcidlink{0000-0002-4644-8818},
Susana C. C. Barros\inst{7,8}\orcidlink{0000-0003-2434-3625},
Fran\c{c}ois Bouchy\inst{9}\orcidlink{0000-0002-7613-393X},
Amadeo Castro-Gonz\'alez\inst{9}\orcidlink{0000-0001-7439-3618},
Olivier D. S. Demangeon\inst{7}\orcidlink{0000-0001-7918-0355},
Xavier Dumusque\inst{9}\orcidlink{0000-0002-9332-2011},
Jonay I. Gonz\'alez Hern\'andez\inst{10,11}\orcidlink{0000-0002-0264-7356},
Baptiste Lavie\inst{9}\orcidlink{0000-0001-8884-9276},
Jorge Lillo-Box\inst{12}\orcidlink{0000-0003-3742-1987},
Christophe Lovis\inst{9}\orcidlink{0000-0001-7120-5837},
Giuseppina Micela\inst{13}\orcidlink{0000-0002-9900-4751},
Enric Palle\inst{10,11}\orcidlink{0000-0003-0987-1593},
Francesco Pepe\inst{9}\orcidlink{0000-0002-9815-773X},
Rafael Rebolo\inst{10,14}\orcidlink{0000-0003-3767-7085},
Jos\'e Rodrigues\inst{7,8,9}\orcidlink{0000-0001-5164-3602},
Nuno C. Santos\inst{7,8}\orcidlink{0000-0003-4422-2919},
S\'ergio G. Sousa\inst{7}\orcidlink{0000-0001-9047-2965},
Alejandro Su\'arez Mascare\~no\inst{10,11}\orcidlink{0000-0002-3814-5323},
St\'ephane Udry\inst{9}\orcidlink{0000-0001-7576-6236},
Mar\'ia R. Zapatero-Osorio\inst{12}\orcidlink{0000-0001-5664-2852}
}

\institute{
\inst{1}Institut Trottier de recherche sur les exoplan\`etes, D\'epartement de Physique, Universit\'e de Montr\'eal, Montr\'eal, Qu\'ebec, Canada\\
\inst{2}Observatoire du Mont-M\'egantic, Qu\'ebec, Canada\\
\inst{3}Univ. Grenoble Alpes, CNRS, IPAG, F-38000 Grenoble, France\\
\inst{4}Department of Physics \& Astronomy, McMaster University, 1280 Main St W, Hamilton, ON, L8S 4L8, Canada\\
\inst{5}Osservatorio Astrofisico di Torino, Via Osservatorio 20, I-10025 Pino Torinese, Italy\\
\inst{6}Center for Space and Habitability, University of Bern, Gesellschaftsstrasse 6, 3012 Bern, Switzerland\\
\inst{7}Instituto de Astrof\'isica e Ci\^encias do Espa\c{c}o, Universidade do Porto, CAUP, Rua das Estrelas, 4150-762 Porto, Portugal\\
\inst{8}Departamento de F\'isica e Astronomia, Faculdade de Ci\^encias, Universidade do Porto, Rua do Campo Alegre, 4169-007 Porto, Portugal\\
\inst{9}Observatoire de Gen\`eve, D\'epartement d’Astronomie, Universit\'e de Gen\`eve, Chemin Pegasi 51, 1290 Versoix, Switzerland\\
\inst{10}Instituto de Astrof\'isica de Canarias (IAC), Calle V\'ia L\'actea s/n, 38205 La Laguna, Tenerife, Spain\\
\inst{11}Departamento de Astrof\'isica, Universidad de La Laguna (ULL), 38206 La Laguna, Tenerife, Spain\\
\inst{12}Centro de Astrobiolog\'ia (CAB), CSIC-INTA, Camino Bajo del Castillo s/n, 28692, Villanueva de la Ca\~nada (Madrid), Spain\\
\inst{13}Osservatorio Astronomico di Palermo, Italy\\
\inst{14}Consejo Superior de Investigaciones Cient\'ificas (CSIC), E-28006 Madrid, Spain\\
\inst{*}\email{avidaan.srivastava@umontreal.ca}
}

   \date{Received XXX; accepted XXX}

\abstract
   {The K2-3 system is one of the most unique planetary systems as it features three transiting sub-Neptunes in decreasing order of radii, spanning the radius valley. In addition to that, the two outer planets are temperate in nature making this system a perfect laboratory for investigating the nature and formation of small potentially habitable worlds. 
   }
   {We present new radial velocity data obtained during the ESPRESSO-GTO programme to add to the archival HARPS, HARPS-N and HIRES data for a total time span of 8 years. This combined dataset significantly improves the masses of the K2-3 planets.}
   {We analyse the archival and ESPRESSO spectroscopic data with the line-by-line algorithm yielding $\sim$60~cm/s precision with ESPRESSO and a $\sim$20\% improvement on previous HARPS and HARPS-N analysis. We coupled this timeseries with a multi-dimensional Gaussian process framework to disentangle the 40.5~days stellar rotation from the Keplerian signals. We tested models using two activity indicators: D2V and H$\alpha$ to ensure the retrieved semi-amplitudes of the planets are not biased by the choice of activity indicator.}
   {The multi-dimensional Gaussian process framework, with either activity indicator, yields the following masses: $M_\text{b}=6.61\pm0.47$\,M$_{\oplus}$; $M_\text{c}=2.23\pm0.68$\,M$_{\oplus}$; $M_\text{d}=2.12\pm1.27$\,M$_{\oplus}$. Compared to previous works, our detection significance of K2-3~b has increased by a factor of two, the mass of K2-3~c has slightly decreased and we now have a marginal detection for K2-3~d. The positions of the planets in the mass-radius diagram suggest either a H/He-rich or volatile-rich composition for both K2-3~b and c, while the nature of K2-3~d is ambiguous. A forward atmospheric model suggests that a single visit with JWST can distinguish between a H$_2$/He-rich atmosphere from a water-world scenario.}
   {This dataset exemplifies the capability of ESPRESSO to detect sub-metre-per-second radial velocity signals for the detection of temperate exoplanets. Additional radial velocity observations are required to improve the detections of K2-3~c and~d, and transit spectroscopy with JWST provides a viable pathway to constrain the atmospheric composition of these worlds.}


\maketitle

\section{Introduction} \label{sec:intro}
With over 6400 exoplanets discovered to date, greater emphasis can now be placed on in-depth study of particularly interesting planetary systems. One of the biggest discoveries in recent years is the radius valley that exists around FGK-type stars~\citep{Fulton2017}. Extending this to the M dwarf sample suggests that late-type stars show a unimodal distribution in planetary radii~\citep{Gillis2026} with early-M stars acting as a transition zone~\citep{CM2020}. The K2-3 system is distinctive as all three planets orbiting the M0 star span the valley, with K2-3~b residing on the mini-Neptune end of the distribution, K2-3~c in the valley and K2-3~d on the super-Earth end.

These planets systematically decrease in size towards longer orbital periods, showing an `inverted' configuration. While such a configuration has been observed for two neighbouring planets in a system (e.g. TOI-1266:~\citealt{Demory2020-TOI1266}; LHS~1903:~\citealt{Wilson2026-LHS1903}), K2-3 remains the only one around an M dwarf where three planets show this architecture. The presence of two temperate ($T_{\text{eq}}\sim250-400\,$K) planets in K2-3~c and d further emphasises the uniqueness of this system and high interest for follow-up atmospheric composition studies.

All these factors have made K2-3 one of the most observed planetary systems with several radial velocity (RV) campaigns from a multitude of instruments that aim to characterise the masses of the three planets~\citep{Almenara2015,Damasso2018,DL2022,Bonomo2023,Howard2025}. This Letter provides an updated look at the system based on 8 years of archival and new RV data. The structure is as follows: Section~\ref{sec:obs} details the various observations, instruments and RV extraction methodologies used for our analysis, Section~\ref{sec:joint-fit} highlights our data analysis process using multi-dimensional Gaussian process to model the stellar activity, Section~\ref{sec:discussion} dives deeper into interpretation of our results in relation to the composition of the planets and prospects of future observations, finally Section~\ref{sec:conclusion} concludes and summarises this work.

\section{Observations} \label{sec:obs}

\subsection{HARPS}

The High Accuracy Radial velocity Planet Searcher (HARPS,~\citealt{Pepe2002}) observed the K2-3 system from 2015 to 2018. The RV dataset was split in two due to the fibre change in 2015~\citep{LoCurto2015}. The 38 pre-change epochs are denoted as `HARPS03' and the 67 post-change as `HARPS15'. 

The full publicly available HARPS dataset was downloaded from DACE~\citep{DACE2015} and the RVs were recomputed using the Line-By-Line (LBL,~\citealt{Dumusque2018,Artigau2022}) algorithm to improve upon the cross-correlation function derived RVs, as seen for other M dwarfs, e.g. ~\citealt{Cadieux2024, Cadieux2025}. The spectra on DACE were reduced by the ESPRESSO-like \texttt{HARPS DRS-3.3.6} pipeline~\citep{Pepe2021} that was optimised for HARPS following a similar approach as for HARPS-N~\citep{Dumusque2021}. Table~\ref{tab:rv_summary} showcases the median RV uncertainty ($\bar\sigma_{RV}$) and the root-mean-square values for each of the datasets used in our analysis. Compared to the template matching algorithm HARPS-TERRA~\citep{TERRA2012} used in~\cite{Damasso2018}, LBL sees a decrease in $\bar \sigma_{RV}$ of $\sim$20\%. Three datapoints were removed due to being low signal-to-noise ratio (S/N) outliers ($>2\bar\sigma_{RV}$).

\begin{figure*}
    \centering
    \includegraphics[width=1\linewidth]{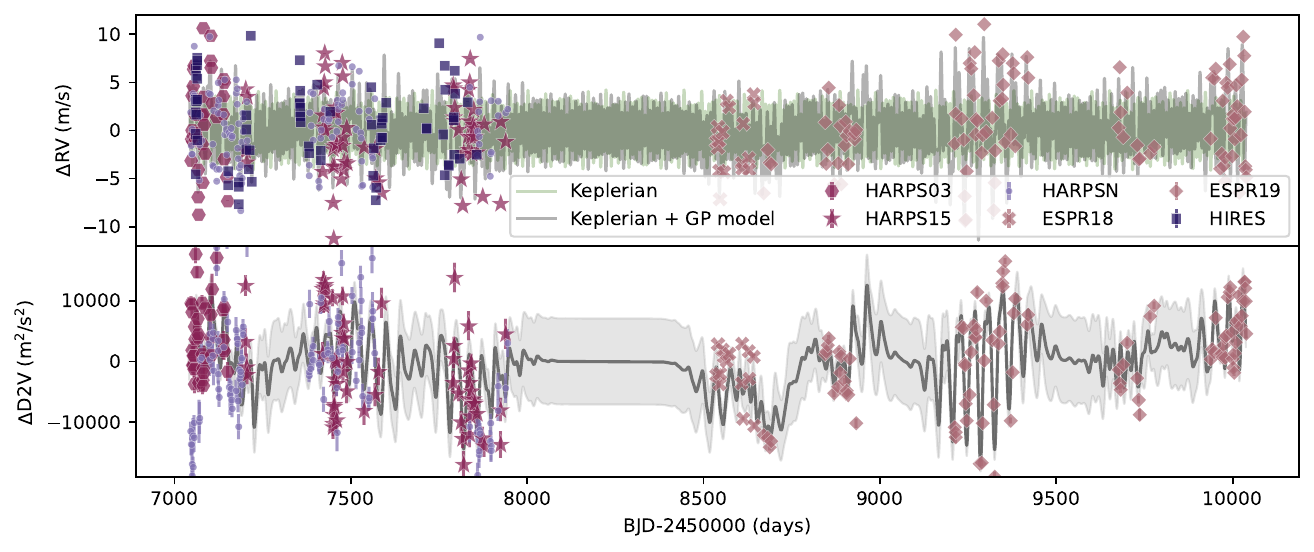}
    \caption{Radial velocity timeseries of K2-3 used in the analysis. Here the second dimension for the multi-dimensional quasi-periodic gaussian process kernes used the D2V activity indicator to disentangle the stellar activity from the Keplerian signals. The Keplerian models for all three planets are assumed to be circular orbits for this figure (Model 1 in Table~\ref{tab:full_model_summary}). The various instruments used in the analysis are colour coded.}
    \label{fig:full_timeseries_d2v}
\end{figure*}

\subsection{HARPS-N}

The High Accuracy Radial velocity Planet Searcher-North (HARPS-N,~\citealt{Cosentino2012}) observed K2-3 from 2015 to 2017. The 128 spectra were once again obtained from DACE, reduced by the similar \texttt{HARPN DRS-3.0.1} data reduction pipeline~\citep{Dumusque2021}. The LBL extracted RVs see a similar improvement in $\bar\sigma_{RV}$ compared to HARPS-TERRA. A total of 11 epochs were discarded due to the low S/N using a similar constraint as for the HARPS dataset. 

\subsection{HIRES}

A similar RV monitoring campaign was launched in the northern hemisphere through the High Resolution Echelle Spectrometer (HIRES,~\citealt{Vogt1999}) between 2015 and 2017. This K2-3 dataset, first published in~\cite{Kosiarek2019}, was used in conjunction with HARPS and HARPS-N in subsequent studies pertaining to this system~\citep{DL2022,Bonomo2023,Howard2025}. Since the wavelength calibrated spectra are not publicly available, we simply use the published RVs as they are on-par with the LBL reduced HARPS and HARPS-N.

\subsection{ESPRESSO}

The Echelle SPectrograph for Rocky Exoplanets and Stable Spectroscopic Observations (ESPRESSO,~\citealt{Pepe2021}) was also used for RV monitoring of the K2-3 system planets from 2019 to 2023. These observations were conducted as part of the Guaranteed Time Observations (GTO) programme, totalling 126 spectra. In 2019 the ESPRESSO fibre-link was upgraded~\citep{Pepe2021}. Thus, similar to HARPS, we denote the 19 ESPRESSO epochs pre-upgrade as `ESPR18' and the 107 post-upgrade as `ESPR19' (8 low-S/N rejections). This data was reduced using \texttt{ESPR DRS-3.3.10}. Per Table~\ref{tab:rv_summary} the LBL retrieved ESPRESSO RVs have the highest quality (by factor of three) compared to the other instruments, making these observations crucial in the RV analysis. 

\section{Radial velocity analysis} \label{sec:final_analysis}

\subsection{Stellar activity}

K2-3 is known to be an active star with an already observed rotation period of $\sim$40\,days based on the K2 photometry and the H$\alpha$ activity indicator~\citep{Damasso2018}. As part of the standard LBL-reduced RV timeseries we also obtain an additional activity indicator D2V~\citep{Zechmeister2018-SERVAL,Artigau2022}, which is the second derivative of the spectrum that acts as a proxy for the full-width half-maximum. 

In previous analyses~\citep{Damasso2018,Bonomo2023,Howard2025} a quasi-periodic Gaussian process (QPGP) kernel was used in addition to the planetary Keplerian models to fit the RV dataset. The QPGP kernel is as follows:
\begin{equation}
    k\,(\tau)\,=\,\text{exp}\Bigg[ -\frac{\text{sin}^2(\pi\tau/P_{\text{rot}})}{2\lambda^2_{\text{P}}} - \frac{\tau^2}{2\lambda^2_{\text{e}}}\Bigg]
\end{equation}

\noindent where $\tau$ is the time lag between two consecutive timestamps in the dataset, P$_{\text{rot}}$ is the periodic signal of the GP attributed to the rotation of the star, $\lambda_{\text{P}}$ is the inverse of the harmonic complexity and $\lambda_{\text{e}}$ is the evolution timescale. This formalism is available by default within the \texttt{Pyaneti} package~\citep{Barragan2019,Barragan2022} which we use for our final RV analysis. 

In order to disentangle the stellar activity from the planetary periods and the respective harmonics, we employ a multi-dimensional GP (MGP) approach. Here, we simultaneously fit the same GP model in at least two different dimensions: the first one is the RV timeseries and the second (and every consequent dimension) is an activity indicator timeseries. By doing this, the GP is constrained by the activity signal and is prevented from over-fitting the RV timeseries, thus preserving the Keplerian signals of the three planets. This technique has been used in past publications~\citep[e.g.][]{Hobson2025,Stefanov2025,Barragan2026} and follows the formalism introduced in~\cite{Rajpaul2015}
\begin{align}
    \Delta \text{RV}   &= A_0\, G(t) + A_1\, G'(t) + K_p + \sigma_{w,1}(t) \tag{Dim. 1} \\
    \Delta \text{Ind1} &= A_2\, G(t) + \sigma_{w,2}(t)\tag{Dim. 2}
\end{align}

\noindent where $\Delta \text{RV}$ is the RV timeseries in the first dimension, $A_n$ are constant coefficients, $G(t)$ is the GP term and $G'(t)$ its derivative, $K_p$ is the Keplerian signals of the planets, `Ind1' is the photometry-like activity indicator and $\sigma_w$ are the white noise components. Any number of activity indicators can be added as extra dimensions, at the expense of a more computationally expensive complex model. We adopt the simplest scenario of a 2D MGP model, using only one activity indicator in conjunction with the RVs. To ensure our results are not biased by the choice of the activity indicator we test both the D2V and H$\alpha$ index timeseries, extracted using the \texttt{ACTIN2}~\citep{ACTIN2} package, in our 2D MGP model. If the results of the two analyses are consistent, we are confident that the 2D MGP framework is adequate for this analysis as both the indicator timeseries will have probed the same stellar modulation.

\subsection{Radial velocity and activity joint fit} \label{sec:joint-fit}

We use \texttt{Pyaneti} and its in-built MGP framework for our joint RV-activity models. Since~\cite{DL2022} provides well constrained radii and ephemeris for the planets, we do not perform a joint RV-transit fit, instead, imposing tight priors on the orbital period and time of inferior conjunction. For each MGP model and activity indicator (D2V and H$\alpha$) pair, we consider two cases: a fixed and a free eccentricity model for K2-3~b. As the detection of K2-3\,c is modest ($\sim3.3\sigma$), and K2-3~d is marginal, we fix the eccentricities of K2-3\,c and d to zero in all cases.

The final planetary parameters and priors for the joint fits are given in Table~\ref{tab:full_model_summary} and the MGP fit using D2V as an activity indicator (Model 1) is shown in Fig.~\ref{fig:full_timeseries_d2v} for reference. Neither of the models with an eccentric orbit for K2-3~b are favoured over the corresponding circular case ($\Delta logZ<1$). The eccentricity obtained from our models is lower than the value reported by~\cite{DL2022}, but consistent within 1$\sigma$ uncertainty. We also obtain a larger semi-amplitude (and mass) for K2-3\,b, while improving the detection significance by a factor of two, and a smaller semi-amplitude for K2-3~c, likely because the MGP framework accounts for stellar activity more effectively than the linear regression implemented in \texttt{EXOFASTv2}~\citep{ExofastV2}. Contrary to~\cite{DL2022}, our analysis hints at a marginal detection for K2-3~d, as indicated by the peaked posterior distribution of its RV semi-amplitude  (Fig.~\ref{fig:corner_Kp}). We therefore report in Table~\ref{tab:planet_summary} the 68\% credible interval for its mass, together with the $3\sigma$ upper limit in parentheses, since a non-detection cannot be excluded. 
    
A simulation involving propagation of the activity signal using fixed values for the GP hyperparameters and randomly sampling similar quality ESPRESSO RVs reveals that approximately 100 additional measurements would be required to constrain the mass of K2-3~d at a 3$\sigma$ level. All models using various activity indicators yield similar RV semi-amplitudes for all planets, consistent within 1$\sigma$ (Table~\ref{tab:full_model_summary}). Since none of the models are  statistically favoured, we report the fitted planetary parameters in Table~\ref{tab:planet_summary} obtained via Gaussian resampling of the posteriors from all four models, providing an average estimate that accounts for the choice of activity indicator.

\begin{figure}
    \centering
    \includegraphics[width=1\linewidth]{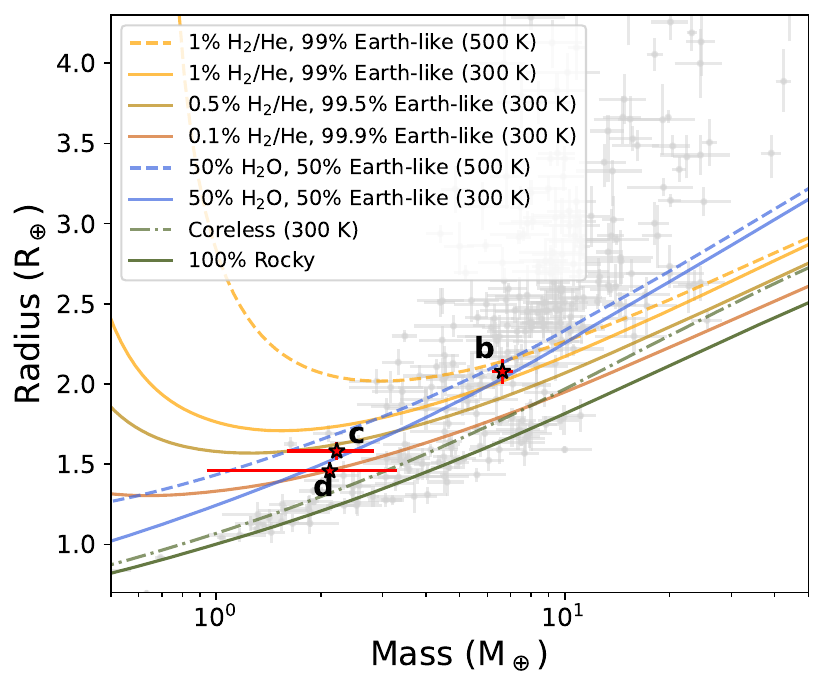}
    \caption{K2-3 planets in mass-radius diagram. The composition lines are taken from the publicly available~\cite{Skinner2026} library. Since the $T_\text{eq}$ for all three planets spans roughly from 300\,k to 500\,K, we present the models for those temperatures.}
    \label{fig:K2-3_MR}
\end{figure}

\section{Discussion} \label{sec:discussion}

\subsection{Mass-Radius diagram} \label{sec:composition}

The K2-3 system is unique in its architecture as it is currently the only known planetary system around an M dwarf star where the three planets' radii systematically decreases as orbital period increases. This inverted configuration coupled with the fact that all three planets span the radius valley make it an interesting system from a planet formation and evolution standpoint. Since early M-type stars show a shallower valley~\citep{CM2020}, the K2-3 planets are intriguing cases for studying the characteristics of the radius valley in a transition zone. 

Placing the planets in the mass-radius diagram (Fig.~\ref{fig:K2-3_MR}) with composition curves taken from~\cite{Skinner2026}, clearly shows that K2-3~b is a sub-Neptune. The well constrained mass and radius suggests a volatile-rich composition, with the solution being degenerate between a water-world or a H$_2$/He atmosphere. Similarly, K2-3~c is compatible with a volatile-rich composition, given its relatively small scaled bulk density~\citep{LuquePalle2022} of $0.39\pm0.13$, making it a strong temperate water-world candidate lying within the radius valley alongside TOI-1452~b~\citep{Cadieux2022} and  LHS~1140~b~\citep{Cadieux2024b,Damiano2024}. K2-3~c joins the broader population of low-density, weakly irradiated ($T_{\mathrm{eq}} \lesssim 500\,\mathrm{K}$) super-Earths identified by~\citet{CastroGonzalez2023}, for which water-rich interiors have been proposed \citep{LuquePalle2022,PiauletGhorayeb2024}. L~98-59~d, with a similar radius but higher equilibrium temperature, hosts a sulphur-rich secondary atmosphere \citep{Gressier2024}, highlighting the need for transit and eclipse observations to establish the nature of K2-3~c. K2-3~d has a more ambiguous composition due to its large mass uncertainty, with a water-world, H$_2$/He-rich and coreless (100\%\,MgSiO$_3$;~\citealt{Coreless2008}) composition all remaining plausible within 1$\sigma$. Its nature therefore remains unconstrained.
Detailed interior modelling informed by stellar refractory abundance ratios (Fe, Mg and Si) measured from infrared spectra, e.g., with NIRPS~\citep{Bouchy2025}~\cite[][]{Cadieux2024,Weisserman2026,Srivastava2026} could better constrain the compositions of all three planets.

\subsection{Prospects for JWST observations} \label{sec:jwst}

To assess the prospects for atmospheric characterisation with JWST~\citep{Gardner2006}, we generated transmission spectra for the three K2-3 planets using the updated system parameters derived in this work. We used \texttt{POSEIDON}~\citep{Poseidon1,Poseidon2} to model two cloud-free atmospheric scenarios: a solar-composition H$_2$/He atmosphere in thermochemical equilibrium with \texttt{FastChem}~\citep{fastchem1,fastchem2}, assuming C/O\,=\,0.55, and an H$_2$O-dominated atmosphere composed of 99.8\% H$_2$O, 0.1\% H$_2$, and 996 ppm He, following the water-rich scenario considered by~\cite{DL2022}. The atmospheres were assumed to be isothermal at the equilibrium temperature of each planet, with the planetary radius defined at 10 bar. We propagated the forward models through PandExo~\citep{Pandexo2017} and simulated one transit with NIRISS/SOSS~\citep{NIRISS2023} and one with NIRSpec/G395H~\citep{NIRSpec2022}. As illustrated in Fig.~\ref{fig:jwst_simu}, the low-mean-molecular-weight H$_2$/He atmospheres produce broad spectral modulations that contrast with the comparatively muted spectra of the H$_2$O-dominated models. Under the photon-noise-limited assumptions adopted here, with no additional systematic noise floor, one transit in each configuration should already distinguish these two broad atmospheric scenarios for all three planets. These forecasts are intentionally illustrative, as clouds or hazes, stellar contamination, and correlated instrumental noise could reduce the observable differences. 

\section{Summary and conclusion} \label{sec:conclusion}
 
The addition of the ESPRESSO dataset has greatly improved the masses and orbital parameters of the K2-3 planets. Our MGP framework has disentangled the stellar rotation period from the Keplerian signals in the RV timeseries using both D2V and H$\alpha$ as activity indicators and obtained consistent results. The models do not strongly favour an eccentric orbit model for K2-3~b over a non-eccentric one. K2-3~b now has a larger mass, K2-3~c has a slightly lower mass and K2-3~d now has a marginal detection, albeit with large uncertainties on the mass, compared to previous publications. The mass-radius diagram suggests a volatile-rich composition for all three planets in the system, however the solutions remain degenerate for the exact chemical makeup of the potential atmospheres of these planets. These planets offer a unique environment for studying the formation and evolution mechanisms that shape temperate planets in the radius valley.

\bibliography{k2-3paper}{}

@article{PiauletGhorayeb2024,
  author  = {Piaulet-Ghorayeb, Caroline and Benneke, Bj{\"o}rn and
             Radica, Michael and Raul, Eshan and Coulombe, Louis-Philippe
             and others},
  title   = {{JWST/NIRISS Reveals the Water-rich ``Steam World''
             Atmosphere of GJ 9827 d}},
  journal = {The Astrophysical Journal Letters},
  volume  = {974},
  number  = {1},
  pages   = {L10},
  year    = {2024}
}

@ARTICLE{Weisserman2026,
       author = {{Weisserman}, Drew and {Gromek}, Nicole and {Cloutier}, Ryan and {Bali}, Komal and {Cadieux}, Charles and {Plotnykov}, Mykhaylo and {L'Heureux}, Alexandrine and {Srivastava}, Avidaan and {Carmona}, Andres and {Frensch}, Yolanda G.~C. and {Artigau}, {\'E}tienne and {Baron}, Fr{\'e}d{\'e}rique and {Barros}, Susana C.~C. and {Benneke}, Bj{\"o}rn and {Bonfils}, Xavier and {Bouchy}, Fran{\c{c}}ois and {Bryan}, Marta and {Cook}, Neil J. and {Cowan}, Nicolas B. and {Cristo}, Eduardo and {Delfosse}, Xavier and {Doyon}, Ren{\'e} and {Dumusque}, Xavier and {Ehrenreich}, David and {Gonz{\'a}lez Hern{\'a}ndez}, Jonay I. and {Lafreni{\`e}re}, David and {de Castro Le{\~a}o}, Izan and {Lovis}, Christophe and {Malo}, Lison and {Canto Martins}, Bruno L. and {Su{\'a}rez Mascare{\~n}o}, Alejandro and {De Medeiros}, Jose Renan and {Melo}, Claudio and {Mignon}, Lucile and {Mordasini}, Christoph and {Pepe}, Francesco and {Rebolo}, Rafael and {Rowe}, Jason and {Santos}, Nuno C. and {S{\'e}gransan}, Damien and {Udry}, St{\'e}phane and {Valencia}, Diana and {Wade}, Gregg and {Aguiar}, Jos{\'e} Luan A. and {Allart}, Romain and {Bazinet}, Luc and {Delisle}, Jean-Baptiste and {B{\'e}langer}, Flavie and {Blackman}, Joshua and {Bourrier}, Vincent and {Branco}, Pedro and {Bruniquel}, Vincent and {Carteret}, Yann and {Cointepas}, Marion and {Darveau-Bernier}, Antoine and {Dauplaise}, Laurie and {Delgado-Mena}, Elisa and {Dorn}, Caroline and {Doshi}, Dhvani and {Faria}, Jo{\~a}o and {Fontinele}, Dasaev O. and {Forveille}, Thierry and {Gagn{\'e}}, Jonathan and {Genest}, Fr{\'e}d{\'e}ric and {Glover}, Jennifer and {de Lima Gomes}, Roseane and {Grieves}, Nolan and {Hobson}, Melissa J. and {Hoeijmakers}, H. Jens and {Jahandar}, Farbod and {Krishnamurthy}, Vigneshwaran and {Lamontagne}, Pierrot and {Larue}, Pierre and {Leath}, Henry and {Lim}, Olivia and {Lipper}, Justin and {Messamah}, Lina and {Messias}, Yuri S. and {Monteiro}, Telmo and {Moranta}, Leslie and {Al Moulla}, Khaled and {Mounzer}, Dany and {Mraz}, Georgia and {Nari}, Nicola and {Nielsen}, Louise D. and {Osborn}, Ares and {Otegi}, Jon and {Parc}, L{\'e}na and {Pelletier}, Stefan and {Pereira}, Olivia and {Piaulet-Ghorayeb}, Caroline and {Rosener}, Riley and {Seidel}, Julia and {Gomes da Silva}, Jo{\~a}o and {Costa Silva}, Ana Rita and {Stefanov}, Atanas K. and {Teixeira}, M{\'a}rcio A. and {Vandal}, Thomas and {Vaulato}, Valentina and {Wardenier}, Joost P. and {Yariv}, Vincent},
        title = "{Super-Earth masses and stellar abundances from NIRPS reveal tentative evidence for water-rich formation around M dwarfs}",
      journal = {\aap},
         year = 2026,
        month = may,
       volume = {709},
          eid = {A165},
        pages = {A165},
archivePrefix = {arXiv},
 primaryClass = {astro-ph.EP},
       adsurl = {https://ui.adsabs.harvard.edu/abs/2026A&A...709A.165W}
}

@ARTICLE{Srivastava2026,
       author = {{Srivastava}, Avidaan and {Doyon}, Ren{\'e} and {Bouchy}, Fran{\c{c}}ois and {Artigau}, {\'E}tienne and {Cadieux}, Charles and {Gromek}, Nicole and {Delgado-Mena}, Elisa and {Messias}, Yuri S. and {Bonfils}, Xavier and {de Lima Gomes}, Roseane and {Barros}, Susana C.~C. and {Benneke}, Bj{\"o}rn and {Bryan}, Marta and {Cloutier}, Ryan and {Cowan}, Nicolas B. and {Cristo}, Eduardo and {Delfosse}, Xavier and {Dumusque}, Xavier and {Ehrenreich}, David and {Gonz{\'a}lez Hern{\'a}ndez}, Jonay I. and {Lafreni{\`e}re}, David and {de Castro Le{\~a}o}, Izan and {Lovis}, Christophe and {Su{\'a}rez Mascare{\~n}o}, Alejandro and {Canto Martins}, Bruno L. and {De Medeiros}, Jose Renan and {Mignon}, Lucile and {Mordasini}, Christoph and {Pepe}, Francesco and {Rebolo}, Rafael and {Rowe}, Jason and {Santos}, Nuno C. and {S{\'e}gransan}, Damien and {Udry}, St{\'e}phane and {Valencia}, Diana and {Wade}, Gregg and {Almenara}, Jose Manuel and {Collins}, Karen A. and {Conti}, Dennis M. and {Dransfield}, George and {Ducrot}, Elsa and {Essack}, Zahra and {Fontinele}, Dasaev O. and {Forveille}, Thierry and {Jafariyazani}, Marziye and {Lamontagne}, Pierrot and {L'Heureux}, Alexandrine and {Al Moulla}, Khaled and {Osborn}, Ares and {Parc}, L{\'e}na and {Rodriguez}, David R. and {Schwarz}, Richard P. and {Scott}, Madison G. and {Shporer}, Avi and {Stefanov}, Atanas K. and {Timmermans}, Mathilde and {Triaud}, Amaury H.~M.~J. and {Wardenier}, Joost P. and {Weisserman}, Drew and {Z{\'u}{\~n}iga-Fern{\'a}ndez}, Sebasti{\'a}n},
        title = "{TOI-4552 b: A new ultra-short-period rocky world revealed by NIRPS and TESS}",
      journal = {\aap},
         year = 2026,
        month = may,
       volume = {709},
          eid = {A73},
        pages = {A73},
archivePrefix = {arXiv},
 primaryClass = {astro-ph.EP},
       adsurl = {https://ui.adsabs.harvard.edu/abs/2026A&A...709A..73S}
}

@ARTICLE{Damasso2018,
       author = {{Damasso}, M. and {Bonomo}, A.~S. and {Astudillo-Defru}, N. and {Bonfils}, X. and {Malavolta}, L. and {Sozzetti}, A. and {Lopez}, E. and {Zeng}, L. and {Haywood}, R.~D. and {Irwin}, J.~M. and {Mortier}, A. and {Vanderburg}, A. and {Maldonado}, J. and {Lanza}, A.~F. and {Affer}, L. and {Almenara}, J.-M. and {Benatti}, S. and {Biazzo}, K. and {Bignamini}, A. and {Borsa}, F. and {Bouchy}, F. and {Buchhave}, L.~A. and {Cameron}, A.~C. and {Carleo}, I. and {Charbonneau}, D. and {Claudi}, R. and {Cosentino}, R. and {Covino}, E. and {Delfosse}, X. and {Desidera}, S. and {Di Fabrizio}, L. and {Dressing}, C. and {Esposito}, M. and {Fares}, R. and {Figueira}, P. and {Fiorenzano}, A.~F.~M. and {Forveille}, T. and {Giacobbe}, P. and {Gonz{\'a}lez-{\'A}lvarez}, E. and {Gratton}, R. and {Harutyunyan}, A. and {Johnson}, J. Asher and {Latham}, D.~W. and {Leto}, G. and {Lopez-Morales}, M. and {Lovis}, C. and {Maggio}, A. and {Mancini}, L. and {Masiero}, S. and {Mayor}, M. and {Micela}, G. and {Molinari}, E. and {Motalebi}, F. and {Murgas}, F. and {Nascimbeni}, V. and {Pagano}, I. and {Pepe}, F. and {Phillips}, D.~F. and {Piotto}, G. and {Poretti}, E. and {Rainer}, M. and {Rice}, K. and {Santos}, N.~C. and {Sasselov}, D. and {Scandariato}, G. and {S{\'e}gransan}, D. and {Smareglia}, R. and {Udry}, S. and {Watson}, C. and {W{\"u}nsche}, A.},
        title = "{Eyes on K2-3: A system of three likely sub-Neptunes characterized with HARPS-N and HARPS}",
      journal = {\aap},
         year = 2018,
        month = jul,
       volume = {615},
          eid = {A69},
        pages = {A69},
archivePrefix = {arXiv},
 primaryClass = {astro-ph.EP},
       adsurl = {https://ui.adsabs.harvard.edu/abs/2018A&A...615A..69D}
}

@ARTICLE{LoCurto2015,
       author = {{Lo Curto}, G. and {Pepe}, F. and {Avila}, G. and {Boffin}, H. and {Bovay}, S. and {Chazelas}, B. and {Coffinet}, A. and {Fleury}, M. and {Hughes}, I. and {Lovis}, C. and {Maire}, C. and {Manescau}, A. and {Pasquini}, L. and {Rihs}, S. and {Sinclaire}, P. and {Udry}, S.},
        title = "{HARPS Gets New Fibres After 12 Years of Operations}",
      journal = {The Messenger},
         year = 2015,
        month = dec,
       volume = {162},
        pages = {9-15},
       adsurl = {https://ui.adsabs.harvard.edu/abs/2015Msngr.162....9L}
}

@ARTICLE{Pepe2002,
       author = {{Pepe}, F. and {Mayor}, M. and {Rupprecht}, G. and {Avila}, G. and {Ballester}, P. and {Beckers}, J.-L. and {Benz}, W. and {Bertaux}, J.-L. and {Bouchy}, F. and {Buzzoni}, B. and {Cavadore}, C. and {Deiries}, S. and {Dekker}, H. and {Delabre}, B. and {D'Odorico}, S. and {Eckert}, W. and {Fischer}, J. and {Fleury}, M. and {George}, M. and {Gilliotte}, A. and {Gojak}, D. and {Guzman}, J.-C. and {Koch}, F. and {Kohler}, D. and {Kotzlowski}, H. and {Lacroix}, D. and {Le Merrer}, J. and {Lizon}, J.-L. and {Lo Curto}, G. and {Longinotti}, A. and {Megevand}, D. and {Pasquini}, L. and {Petitpas}, P. and {Pichard}, M. and {Queloz}, D. and {Reyes}, J. and {Richaud}, P. and {Sivan}, J.-P. and {Sosnowska}, D. and {Soto}, R. and {Udry}, S. and {Ureta}, E. and {van Kesteren}, A. and {Weber}, L. and {Weilenmann}, U. and {Wicenec}, A. and {Wieland}, G. and {Christensen-Dalsgaard}, J. and {Dravins}, D. and {Hatzes}, A. and {K{\"u}rster}, M. and {Paresce}, F. and {Penny}, A.},
        title = "{HARPS: ESO's coming planet searcher. Chasing exoplanets with the La Silla 3.6-m telescope}",
      journal = {The Messenger},
         year = 2002,
        month = dec,
       volume = {110},
        pages = {9-14},
       adsurl = {https://ui.adsabs.harvard.edu/abs/2002Msngr.110....9P}
}

@ARTICLE{Artigau2022,
       author = {{Artigau}, {\'E}tienne and {Cadieux}, Charles and {Cook}, Neil J. and {Doyon}, Ren{\'e} and {Vandal}, Thomas and {Donati}, Jean-Fran{\c{c}}ois and {Moutou}, Claire and {Delfosse}, Xavier and {Fouqu{\'e}}, Pascal and {Martioli}, Eder and {Bouchy}, Fran{\c{c}}ois and {Parsons}, Jasmine and {Carmona}, Andres and {Dumusque}, Xavier and {Astudillo-Defru}, Nicola and {Bonfils}, Xavier and {Mignon}, Lucille},
        title = "{Line-by-line Velocity Measurements: an Outlier-resistant Method for Precision Velocimetry}",
      journal = {\aj},
         year = 2022,
        month = sep,
       volume = {164},
       number = {3},
          eid = {84},
        pages = {84},
archivePrefix = {arXiv},
 primaryClass = {astro-ph.IM},
       adsurl = {https://ui.adsabs.harvard.edu/abs/2022AJ....164...84A}
}

@ARTICLE{Pepe2021,
       author = {{Pepe}, F. and {Cristiani}, S. and {Rebolo}, R. and {Santos}, N.~C. and {Dekker}, H. and {Cabral}, A. and {Di Marcantonio}, P. and {Figueira}, P. and {Lo Curto}, G. and {Lovis}, C. and {Mayor}, M. and {M{\'e}gevand}, D. and {Molaro}, P. and {Riva}, M. and {Zapatero Osorio}, M.~R. and {Amate}, M. and {Manescau}, A. and {Pasquini}, L. and {Zerbi}, F.~M. and {Adibekyan}, V. and {Abreu}, M. and {Affolter}, M. and {Alibert}, Y. and {Aliverti}, M. and {Allart}, R. and {Allende Prieto}, C. and {{\'A}lvarez}, D. and {Alves}, D. and {Avila}, G. and {Baldini}, V. and {Bandy}, T. and {Barros}, S.~C.~C. and {Benz}, W. and {Bianco}, A. and {Borsa}, F. and {Bourrier}, V. and {Bouchy}, F. and {Broeg}, C. and {Calderone}, G. and {Cirami}, R. and {Coelho}, J. and {Conconi}, P. and {Coretti}, I. and {Cumani}, C. and {Cupani}, G. and {D'Odorico}, V. and {Damasso}, M. and {Deiries}, S. and {Delabre}, B. and {Demangeon}, O.~D.~S. and {Dumusque}, X. and {Ehrenreich}, D. and {Faria}, J.~P. and {Fragoso}, A. and {Genolet}, L. and {Genoni}, M. and {G{\'e}nova Santos}, R. and {Gonz{\'a}lez Hern{\'a}ndez}, J.~I. and {Hughes}, I. and {Iwert}, O. and {Kerber}, F. and {Knudstrup}, J. and {Landoni}, M. and {Lavie}, B. and {Lillo-Box}, J. and {Lizon}, J.-L. and {Maire}, C. and {Martins}, C.~J.~A.~P. and {Mehner}, A. and {Micela}, G. and {Modigliani}, A. and {Monteiro}, M.~A. and {Monteiro}, M.~J.~P.~F.~G. and {Moschetti}, M. and {Murphy}, M.~T. and {Nunes}, N. and {Oggioni}, L. and {Oliveira}, A. and {Oshagh}, M. and {Pall{\'e}}, E. and {Pariani}, G. and {Poretti}, E. and {Rasilla}, J.~L. and {Rebord{\~a}o}, J. and {Redaelli}, E.~M. and {Santana Tschudi}, S. and {Santin}, P. and {Santos}, P. and {S{\'e}gransan}, D. and {Schmidt}, T.~M. and {Segovia}, A. and {Sosnowska}, D. and {Sozzetti}, A. and {Sousa}, S.~G. and {Span{\`o}}, P. and {Su{\'a}rez Mascare{\~n}o}, A. and {Tabernero}, H. and {Tenegi}, F. and {Udry}, S. and {Zanutta}, A.},
        title = "{ESPRESSO at VLT. On-sky performance and first results}",
      journal = {\aap},
         year = 2021,
        month = jan,
       volume = {645},
          eid = {A96},
        pages = {A96},
archivePrefix = {arXiv},
 primaryClass = {astro-ph.IM},
       adsurl = {https://ui.adsabs.harvard.edu/abs/2021A&A...645A..96P}
}

@INPROCEEDINGS{Cosentino2012,
       author = {{Cosentino}, Rosario and {Lovis}, Christophe and {Pepe}, Francesco and {Collier Cameron}, Andrew and {Latham}, David W. and {Molinari}, Emilio and {Udry}, Stephane and {Bezawada}, Naidu and {Black}, Martin and {Born}, Andy and {Buchschacher}, Nicolas and {Charbonneau}, Dave and {Figueira}, Pedro and {Fleury}, Michel and {Galli}, Alberto and {Gallie}, Angus and {Gao}, Xiaofeng and {Ghedina}, Adriano and {Gonzalez}, Carlos and {Gonzalez}, Manuel and {Guerra}, Jose and {Henry}, David and {Horne}, Keith and {Hughes}, Ian and {Kelly}, Dennis and {Lodi}, Marcello and {Lunney}, David and {Maire}, Charles and {Mayor}, Michel and {Micela}, Giusi and {Ordway}, Mark P. and {Peacock}, John and {Phillips}, David and {Piotto}, Giampaolo and {Pollacco}, Don and {Queloz}, Didier and {Rice}, Ken and {Riverol}, Carlos and {Riverol}, Luis and {San Juan}, Jose and {Sasselov}, Dimitar and {Segransan}, Damien and {Sozzetti}, Alessandro and {Sosnowska}, Danuta and {Stobie}, Brian and {Szentgyorgyi}, Andrew and {Vick}, Andy and {Weber}, Luc},
        title = "{Harps-N: the new planet hunter at TNG}",
    booktitle = {Ground-based and Airborne Instrumentation for Astronomy IV},
         year = 2012,
       editor = {{McLean}, Ian S. and {Ramsay}, Suzanne K. and {Takami}, Hideki},
       series = {Society of Photo-Optical Instrumentation Engineers (SPIE) Conference Series},
       volume = {8446},
        month = sep,
          eid = {84461V},
        pages = {84461V},
       adsurl = {https://ui.adsabs.harvard.edu/abs/2012SPIE.8446E..1VC}
}

@INPROCEEDINGS{DACE2015,
       author = {{Buchschacher}, N. and {S{\'e}gransan}, D. and {Udry}, S. and {D{\'\i}az}, R.},
        title = "{Data and Analysis Center for Exoplanets}",
    booktitle = {Astronomical Data Analysis Software and Systems XXIV (ADASS XXIV)},
         year = 2015,
       editor = {{Taylor}, A.~R. and {Rosolowsky}, E.},
       series = {Astronomical Society of the Pacific Conference Series},
       volume = {495},
        month = sep,
        pages = {7},
       adsurl = {https://ui.adsabs.harvard.edu/abs/2015ASPC..495....7B}
}

@ARTICLE{Kosiarek2019,
       author = {{Kosiarek}, Molly R. and {Crossfield}, Ian J.~M. and {Hardegree-Ullman}, Kevin K. and {Livingston}, John H. and {Benneke}, Bj{\"o}rn and {Henry}, Gregory W. and {Howard}, Ward S. and {Berardo}, David and {Blunt}, Sarah and {Fulton}, Benjamin J. and {Hirsch}, Lea A. and {Howard}, Andrew W. and {Isaacson}, Howard and {Petigura}, Erik A. and {Sinukoff}, Evan and {Weiss}, Lauren and {Bonfils}, X. and {Dressing}, Courtney D. and {Knutson}, Heather A. and {Schlieder}, Joshua E. and {Werner}, Michael and {Gorjian}, Varoujan and {Krick}, Jessica and {Morales}, Farisa Y. and {Astudillo-Defru}, Nicola and {Almenara}, J.-M. and {Delfosse}, X. and {Forveille}, T. and {Lovis}, C. and {Mayor}, M. and {Murgas}, F. and {Pepe}, F. and {Santos}, N.~C. and {Udry}, S. and {Corbett}, H.~T. and {Fors}, Octavi and {Law}, Nicholas M. and {Ratzloff}, Jeffrey K. and {del Ser}, Daniel},
        title = "{Bright Opportunities for Atmospheric Characterization of Small Planets: Masses and Radii of K2-3 b, c, and d and GJ3470 b from Radial Velocity Measurements and Spitzer Transits}",
      journal = {\aj},
         year = 2019,
        month = mar,
       volume = {157},
       number = {3},
          eid = {97},
        pages = {97},
archivePrefix = {arXiv},
 primaryClass = {astro-ph.EP},
       adsurl = {https://ui.adsabs.harvard.edu/abs/2019AJ....157...97K}
}

@ARTICLE{DL2022,
       author = {{Diamond-Lowe}, Hannah and {Kreidberg}, Laura and {Harman}, C.~E. and {Kempton}, Eliza M.-R. and {Rogers}, Leslie A. and {Joyce}, Simon R.~G. and {Eastman}, Jason D. and {King}, George W. and {Kopparapu}, Ravi and {Youngblood}, Allison and {Kosiarek}, Molly R. and {Livingston}, John H. and {Hardegree-Ullman}, Kevin K. and {Crossfield}, Ian J.~M.},
        title = "{The K2-3 System Revisited: Testing Photoevaporation and Core-powered Mass Loss with Three Small Planets Spanning the Radius Valley}",
      journal = {\aj},
         year = 2022,
        month = nov,
       volume = {164},
       number = {5},
          eid = {172},
        pages = {172},
archivePrefix = {arXiv},
 primaryClass = {astro-ph.EP},
       adsurl = {https://ui.adsabs.harvard.edu/abs/2022AJ....164..172D}
}

@ARTICLE{Bonomo2023,
       author = {{Bonomo}, A.~S. and {Dumusque}, X. and {Massa}, A. and {Mortier}, A. and {Bongiolatti}, R. and {Malavolta}, L. and {Sozzetti}, A. and {Buchhave}, L.~A. and {Damasso}, M. and {Haywood}, R.~D. and {Morbidelli}, A. and {Latham}, D.~W. and {Molinari}, E. and {Pepe}, F. and {Poretti}, E. and {Udry}, S. and {Affer}, L. and {Boschin}, W. and {Charbonneau}, D. and {Cosentino}, R. and {Cretignier}, M. and {Ghedina}, A. and {Lega}, E. and {L{\'o}pez-Morales}, M. and {Margini}, M. and {Mart{\'\i}nez Fiorenzano}, A.~F. and {Mayor}, M. and {Micela}, G. and {Pedani}, M. and {Pinamonti}, M. and {Rice}, K. and {Sasselov}, D. and {Tronsgaard}, R. and {Vanderburg}, A.},
        title = "{Cold Jupiters and improved masses in 38 Kepler and K2 small planet systems from 3661 HARPS-N radial velocities. No excess of cold Jupiters in small planet systems}",
      journal = {\aap},
         year = 2023,
        month = sep,
       volume = {677},
          eid = {A33},
        pages = {A33},
archivePrefix = {arXiv},
 primaryClass = {astro-ph.EP},
       adsurl = {https://ui.adsabs.harvard.edu/abs/2023A&A...677A..33B}
}

@ARTICLE{Howard2025,
       author = {{Howard}, Andrew W. and {Sinukoff}, Evan and {Blunt}, Sarah and {Petigura}, Erik A. and {Crossfield}, Ian J.~M. and {Isaacson}, Howard and {Kosiarek}, Molly and {Rubenzahl}, Ryan A. and {Brewer}, John M. and {Fulton}, Benjamin J. and {Dressing}, Courtney D. and {Hirsch}, Lea A. and {Knutson}, Heather and {Livingston}, John H. and {Mills}, Sean M. and {Roy}, Arpita and {Weiss}, Lauren M. and {Benneke}, Bjorn and {Ciardi}, David R. and {Christiansen}, Jessie L. and {Cochran}, William D. and {Crepp}, Justin R. and {Gonzales}, Erica and {Hansen}, Brad M.~S. and {Hardegree-Ullman}, Kevin and {Howell}, Steve B. and {L{\'e}pine}, S{\'e}bastien and {Martinez}, Arturo O. and {Rogers}, Leslie A. and {Schlieder}, Joshua E. and {Werner}, Michael and {Polanski}, Alex S. and {Angelo}, Isabel and {Beard}, Corey and {Behmard}, Aida and {Bouma}, Luke G. and {Brinkman}, Casey L. and {Chontos}, Ashley and {Dai}, Fei and {Dalba}, Paul A. and {Giacalone}, Steven and {Grunblatt}, Samuel K. and {Hill}, Michelle L. and {Kane}, Stephen R. and {Lubin}, Jack and {Mayo}, Andrew W. and {Mocnik}, Teo and {Murphy}, Joseph M. Akana and {Rice}, Malena and {Rosenthal}, Lee J. and {Tyler}, Dakotah and {Van Zandt}, Judah and {Yee}, Samuel W.},
        title = "{Planet Masses, Radii, and Orbits from NASA's K2 Mission}",
      journal = {\apjs},
         year = 2025,
        month = jun,
       volume = {278},
       number = {2},
          eid = {52},
        pages = {52},
archivePrefix = {arXiv},
 primaryClass = {astro-ph.EP},
       adsurl = {https://ui.adsabs.harvard.edu/abs/2025ApJS..278...52H}
}

@software{Zechmeister2018-GLS,
       author = {{Zechmeister}, Mathias and {K{\"u}rster}, M.},
        title = "{GLS: Generalized Lomb-Scargle periodogram}",
 howpublished = {Astrophysics Source Code Library, record ascl:1807.019},
         year = 2018,
        month = jul,
          eid = {ascl:1807.019},
archivePrefix = {ascl},
       adsurl = {https://ui.adsabs.harvard.edu/abs/2018ascl.soft07019Z}
}

@ARTICLE{Zechmeister2018-SERVAL,
       author = {{Zechmeister}, M. and {Reiners}, A. and {Amado}, P.~J. and {Azzaro}, M. and {Bauer}, F.~F. and {B{\'e}jar}, V.~J.~S. and {Caballero}, J.~A. and {Guenther}, E.~W. and {Hagen}, H.-J. and {Jeffers}, S.~V. and {Kaminski}, A. and {K{\"u}rster}, M. and {Launhardt}, R. and {Montes}, D. and {Morales}, J.~C. and {Quirrenbach}, A. and {Reffert}, S. and {Ribas}, I. and {Seifert}, W. and {Tal-Or}, L. and {Wolthoff}, V.},
        title = "{Spectrum radial velocity analyser (SERVAL). High-precision radial velocities and two alternative spectral indicators}",
      journal = {\aap},
         year = 2018,
        month = jan,
       volume = {609},
          eid = {A12},
        pages = {A12},
archivePrefix = {arXiv},
 primaryClass = {astro-ph.IM},
       adsurl = {https://ui.adsabs.harvard.edu/abs/2018A&A...609A..12Z}
}

@ARTICLE{Barragan2022,
       author = {{Barrag{\'a}n}, Oscar and {Aigrain}, Suzanne and {Rajpaul}, Vinesh M. and {Zicher}, Norbert},
        title = "{PYANETI - II. A multidimensional Gaussian process approach to analysing spectroscopic time-series}",
      journal = {\mnras},
         year = 2022,
        month = jan,
       volume = {509},
       number = {1},
        pages = {866-883},
archivePrefix = {arXiv},
 primaryClass = {astro-ph.EP},
       adsurl = {https://ui.adsabs.harvard.edu/abs/2022MNRAS.509..866B}
}

@ARTICLE{Barragan2019,
       author = {{Barrag{\'a}n}, O. and {Gandolfi}, D. and {Antoniciello}, G.},
        title = "{PYANETI: a fast and powerful software suite for multiplanet radial velocity and transit fitting}",
      journal = {\mnras},
         year = 2019,
        month = jan,
       volume = {482},
       number = {1},
        pages = {1017-1030},
archivePrefix = {arXiv},
 primaryClass = {astro-ph.EP},
       adsurl = {https://ui.adsabs.harvard.edu/abs/2019MNRAS.482.1017B}
}

@ARTICLE{Barragan2026,
       author = {{Barrag{\'a}n}, Oscar and {Mallorqu{\'\i}n}, Manuel and {Fern{\'a}ndez-Fern{\'a}ndez}, Jorge and {Hawthorn}, Faith and {Freckelton}, Alix V. and {Lafarga}, Marina and {Cretignier}, Michael and {Eschen}, Yoshi N.~E. and {Gill}, Samuel and {B{\'e}jar}, V{\'\i}ctor J.~S. and {Lodieu}, Nicolas and {Yu}, Haochuan and {Wilson}, Thomas G. and {Anderson}, David and {Apergis}, Ioannis and {Battley}, Matthew and {Bryant}, Edward M. and {Cort{\'e}s-Zuleta}, P{\'\i}a and {Gillen}, Edward and {Jenkins}, James S. and {Klein}, Baptiste and {McCormac}, James and {Meech}, Annabella and {Meier-Vald{\'e}s}, Erik and {Moyano}, Maximiliano and {Mortier}, Annelies and {Murgas}, Felipe and {Nielsen}, Louise D. and {Saha}, Suman and {Vines}, Jos{\'e} I. and {West}, Richard and {Wheatley}, Peter J. and {Aigrain}, Suzanne},
        title = "{Mass estimates of the young TOI-451 transiting planets: multidimensional Gaussian Process on stellar spectroscopic and photometric signals}",
      journal = {\mnras},
         year = 2026,
        month = feb,
       volume = {546},
       number = {2},
          eid = {stag087},
        pages = {stag087},
archivePrefix = {arXiv},
 primaryClass = {astro-ph.EP},
       adsurl = {https://ui.adsabs.harvard.edu/abs/2026MNRAS.546ag087B}
}

@ARTICLE{Stefanov2025,
       author = {{Stefanov}, A.~K. and {Gonz{\'a}lez Hern{\'a}ndez}, J.~I. and {Su{\'a}rez Mascare{\~n}o}, A. and {Rebolo}, R. and {Nari}, N. and {Mestre}, J.~M. and {Sousa}, S.~G. and {Tabernero}, H.~M. and {Zapatero Osorio}, M.-R. and {Figueira}, P. and {Faria}, J.~P. and {Hobson}, M.~J. and {Silva}, A.~M. and {Castro-Gonz{\'a}lez}, A. and {Santos}, N.~C. and {Sozzetti}, A. and {Pepe}, F. and {Cristiani}, S. and {Lavie}, B. and {Martins}, C.~J.~A.~P.},
        title = "{Stellar-activity analysis of the nearby M dwarf GJ 526: Multi-dimensional Gaussian-process modelling of RV, FWHM, and S-index}",
      journal = {\aap},
         year = 2025,
        month = nov,
       volume = {703},
          eid = {A245},
        pages = {A245},
archivePrefix = {arXiv},
 primaryClass = {astro-ph.SR},
       adsurl = {https://ui.adsabs.harvard.edu/abs/2025A&A...703A.245S}
}

@ARTICLE{Rajpaul2015,
       author = {{Rajpaul}, V. and {Aigrain}, S. and {Osborne}, M.~A. and {Reece}, S. and {Roberts}, S.},
        title = "{A Gaussian process framework for modelling stellar activity signals in radial velocity data}",
      journal = {\mnras},
         year = 2015,
        month = sep,
       volume = {452},
       number = {3},
        pages = {2269-2291},
archivePrefix = {arXiv},
 primaryClass = {astro-ph.EP},
       adsurl = {https://ui.adsabs.harvard.edu/abs/2015MNRAS.452.2269R}
}

@ARTICLE{Kempton2018,
       author = {{Kempton}, Eliza M.-R. and {Bean}, Jacob L. and {Louie}, Dana R. and {Deming}, Drake and {Koll}, Daniel D.~B. and {Mansfield}, Megan and {Christiansen}, Jessie L. and {L{\'o}pez-Morales}, Mercedes and {Swain}, Mark R. and {Zellem}, Robert T. and {Ballard}, Sarah and {Barclay}, Thomas and {Barstow}, Joanna K. and {Batalha}, Natasha E. and {Beatty}, Thomas G. and {Berta-Thompson}, Zach and {Birkby}, Jayne and {Buchhave}, Lars A. and {Charbonneau}, David and {Cowan}, Nicolas B. and {Crossfield}, Ian and {de Val-Borro}, Miguel and {Doyon}, Ren{\'e} and {Dragomir}, Diana and {Gaidos}, Eric and {Heng}, Kevin and {Hu}, Renyu and {Kane}, Stephen R. and {Kreidberg}, Laura and {Mallonn}, Matthias and {Morley}, Caroline V. and {Narita}, Norio and {Nascimbeni}, Valerio and {Pall{\'e}}, Enric and {Quintana}, Elisa V. and {Rauscher}, Emily and {Seager}, Sara and {Shkolnik}, Evgenya L. and {Sing}, David K. and {Sozzetti}, Alessandro and {Stassun}, Keivan G. and {Valenti}, Jeff A. and {von Essen}, Carolina},
        title = "{A Framework for Prioritizing the TESS Planetary Candidates Most Amenable to Atmospheric Characterization}",
      journal = {\pasp},
         year = 2018,
        month = nov,
       volume = {130},
       number = {993},
        pages = {114401},
archivePrefix = {arXiv},
 primaryClass = {astro-ph.EP},
       adsurl = {https://ui.adsabs.harvard.edu/abs/2018PASP..130k4401K}
}

@ARTICLE{Gillis2026,
       author = {{Gillis}, Erik Diego and {Cloutier}, Ryan and {Pass}, Emily K.},
        title = "{TESS Planet Occurrence Rates Reveal the Disappearance of the Radius Valley around Mid-to-late M Dwarfs}",
      journal = {\aj},
         year = 2026,
        month = may,
       volume = {171},
       number = {5},
          eid = {317},
        pages = {317},
archivePrefix = {arXiv},
 primaryClass = {astro-ph.EP},
       adsurl = {https://ui.adsabs.harvard.edu/abs/2026AJ....171..317G}
}

@ARTICLE{CM2020,
       author = {{Cloutier}, Ryan and {Menou}, Kristen},
        title = "{Evolution of the Radius Valley around Low-mass Stars from Kepler and K2}",
      journal = {\aj},
         year = 2020,
        month = may,
       volume = {159},
       number = {5},
          eid = {211},
        pages = {211},
archivePrefix = {arXiv},
 primaryClass = {astro-ph.EP},
       adsurl = {https://ui.adsabs.harvard.edu/abs/2020AJ....159..211C}
}

@ARTICLE{Skinner2026,
       author = {{Skinner}, Bennett Neil and {Pudritz}, Ralph E. and {Cloutier}, Ryan},
        title = "{A validated low-to-intermediate mass planetary interior structure model and new mass─radius relations}",
      journal = {\mnras},
         year = 2026,
        month = aug,
       volume = {550},
       number = {3},
          eid = {stag1076},
        pages = {stag1076},
archivePrefix = {arXiv},
 primaryClass = {astro-ph.EP},
       adsurl = {https://ui.adsabs.harvard.edu/abs/2026MNRAS.550g1076S}
}

@ARTICLE{Cadieux2022,
       author = {{Cadieux}, Charles and {Doyon}, Ren{\'e} and {Plotnykov}, Mykhaylo and {H{\'e}brard}, Guillaume and {Jahandar}, Farbod and {Artigau}, {\'E}tienne and {Valencia}, Diana and {Cook}, Neil J. and {Martioli}, Eder and {Vandal}, Thomas and {Donati}, Jean-Fran{\c{c}}ois and {Cloutier}, Ryan and {Narita}, Norio and {Fukui}, Akihiko and {Hirano}, Teruyuki and {Bouchy}, Fran{\c{c}}ois and {Cowan}, Nicolas B. and {Gonzales}, Erica J. and {Ciardi}, David R. and {Stassun}, Keivan G. and {Arnold}, Luc and {Benneke}, Bj{\"o}rn and {Boisse}, Isabelle and {Bonfils}, Xavier and {Carmona}, Andr{\'e}s and {Cort{\'e}s-Zuleta}, P{\'\i}a and {Delfosse}, Xavier and {Forveille}, Thierry and {Fouqu{\'e}}, Pascal and {Gomes da Silva}, Jo{\~a}o and {Jenkins}, Jon M. and {Kiefer}, Flavien and {K{\'o}sp{\'a}l}, {\'A}gnes and {Lafreni{\`e}re}, David and {Martins}, Jorge H.~C. and {Moutou}, Claire and {do Nascimento}, J.-D. and {Ould-Elhkim}, Merwan and {Pelletier}, Stefan and {Twicken}, Joseph D. and {Bouma}, Luke G. and {Cartwright}, Scott and {Darveau-Bernier}, Antoine and {Grankin}, Konstantin and {Ikoma}, Masahiro and {Kagetani}, Taiki and {Kawauchi}, Kiyoe and {Kodama}, Takanori and {Kotani}, Takayuki and {Latham}, David W. and {Menou}, Kristen and {Ricker}, George and {Seager}, Sara and {Tamura}, Motohide and {Vanderspek}, Roland and {Watanabe}, Noriharu},
        title = "{TOI-1452 b: SPIRou and TESS Reveal a Super-Earth in a Temperate Orbit Transiting an M4 Dwarf}",
      journal = {\aj},
         year = 2022,
        month = sep,
       volume = {164},
       number = {3},
          eid = {96},
        pages = {96},
archivePrefix = {arXiv},
 primaryClass = {astro-ph.EP},
       adsurl = {https://ui.adsabs.harvard.edu/abs/2022AJ....164...96C}
}

@ARTICLE{Cadieux2024,
       author = {{Cadieux}, Charles and {Plotnykov}, Mykhaylo and {Doyon}, Ren{\'e} and {Valencia}, Diana and {Jahandar}, Farbod and {Dang}, Lisa and {Turbet}, Martin and {Fauchez}, Thomas J. and {Cloutier}, Ryan and {Cherubim}, Collin and {Artigau}, {\'E}tienne and {Cook}, Neil J. and {Edwards}, Billy and {Hallatt}, Tim and {Charnay}, Benjamin and {Bouchy}, Fran{\c{c}}ois and {Allart}, Romain and {Mignon}, Lucile and {Baron}, Fr{\'e}d{\'e}rique and {Barros}, Susana C.~C. and {Benneke}, Bj{\"o}rn and {Canto Martins}, B.~L. and {Cowan}, Nicolas B. and {De Medeiros}, J.~R. and {Delfosse}, Xavier and {Delgado-Mena}, Elisa and {Dumusque}, Xavier and {Ehrenreich}, David and {Frensch}, Yolanda G.~C. and {Gonz{\'a}lez Hern{\'a}ndez}, J.~I. and {Hara}, Nathan C. and {Lafreni{\`e}re}, David and {Lo Curto}, Gaspare and {Malo}, Lison and {Melo}, Claudio and {Mounzer}, Dany and {Passeger}, Vera Maria and {Pepe}, Francesco and {Poulin-Girard}, Anne-Sophie and {Santos}, Nuno C. and {Sosnowska}, Danuta and {Su{\'a}rez Mascare{\~n}o}, Alejandro and {Thibault}, Simon and {Vaulato}, Valentina and {Wade}, Gregg A. and {Wildi}, Fran{\c{c}}ois},
        title = "{New Mass and Radius Constraints on the LHS 1140 Planets: LHS 1140 b Is either a Temperate Mini-Neptune or a Water World}",
      journal = {\apjl},
         year = 2024,
        month = jan,
       volume = {960},
       number = {1},
          eid = {L3},
        pages = {L3},
archivePrefix = {arXiv},
 primaryClass = {astro-ph.EP},
       adsurl = {https://ui.adsabs.harvard.edu/abs/2024ApJ...960L...3C}
}

@ARTICLE{Gressier2024,
       author = {{Gressier}, Am{\'e}lie and {Espinoza}, N{\'e}stor and {Allen}, Natalie H. and {Sing}, David K. and {Banerjee}, Agnibha and {Barstow}, Joanna K. and {Valenti}, Jeff A. and {Lewis}, Nikole K. and {Birkmann}, Stephan M. and {Challener}, Ryan C. and {Manjavacas}, Elena and {Alves de Oliveira}, Catarina and {Crouzet}, Nicolas and {Beck}, Tracy. L.},
        title = "{Hints of a Sulfur-rich Atmosphere around the 1.6 R $_{{\ensuremath{\oplus}}}$ Super-Earth L98-59 d from JWST NIRspec G395H Transmission Spectroscopy}",
      journal = {\apjl},
         year = 2024,
        month = nov,
       volume = {975},
       number = {1},
          eid = {L10},
        pages = {L10},
archivePrefix = {arXiv},
 primaryClass = {astro-ph.EP},
       adsurl = {https://ui.adsabs.harvard.edu/abs/2024ApJ...975L..10G}
}

@ARTICLE{Coreless2008,
       author = {{Elkins-Tanton}, Linda T. and {Seager}, Sara},
        title = "{Coreless Terrestrial Exoplanets}",
      journal = {\apj},
         year = 2008,
        month = nov,
       volume = {688},
       number = {1},
        pages = {628-635},
archivePrefix = {arXiv},
 primaryClass = {astro-ph},
       adsurl = {https://ui.adsabs.harvard.edu/abs/2008ApJ...688..628E}
}

@ARTICLE{Gardner2006,
       author = {{Gardner}, Jonathan P. and {Mather}, John C. and {Clampin}, Mark and {Doyon}, Rene and {Greenhouse}, Matthew A. and {Hammel}, Heidi B. and {Hutchings}, John B. and {Jakobsen}, Peter and {Lilly}, Simon J. and {Long}, Knox S. and {Lunine}, Jonathan I. and {McCaughrean}, Mark J. and {Mountain}, Matt and {Nella}, John and {Rieke}, George H. and {Rieke}, Marcia J. and {Rix}, Hans-Walter and {Smith}, Eric P. and {Sonneborn}, George and {Stiavelli}, Massimo and {Stockman}, H.~S. and {Windhorst}, Rogier A. and {Wright}, Gillian S.},
        title = "{The James Webb Space Telescope}",
      journal = {\ssr},
         year = 2006,
        month = apr,
       volume = {123},
       number = {4},
        pages = {485-606},
archivePrefix = {arXiv},
 primaryClass = {astro-ph},
       adsurl = {https://ui.adsabs.harvard.edu/abs/2006SSRv..123..485G}
}

@ARTICLE{Poseidon1,
       author = {{MacDonald}, Ryan J. and {Madhusudhan}, Nikku},
        title = "{HD 209458b in new light: evidence of nitrogen chemistry, patchy clouds and sub-solar water}",
      journal = {\mnras},
         year = 2017,
        month = aug,
       volume = {469},
       number = {2},
        pages = {1979-1996},
archivePrefix = {arXiv},
 primaryClass = {astro-ph.EP},
       adsurl = {https://ui.adsabs.harvard.edu/abs/2017MNRAS.469.1979M}
}

@article{Poseidon2, 
    year = {2023}, 
    publisher = {The Open Journal}, 
    volume = {8}, 
    number = {81}, 
    pages = {4873}, 
    author = {MacDonald, Ryan J.}, 
    title = {POSEIDON: A Multidimensional Atmospheric Retrieval Code for Exoplanet Spectra}, 
    journal = {Journal of Open Source Software} }

@ARTICLE{Pandexo2017,
       author = {{Batalha}, Natasha E. and {Mandell}, Avi and {Pontoppidan}, Klaus and {Stevenson}, Kevin B. and {Lewis}, Nikole K. and {Kalirai}, Jason and {Earl}, Nick and {Greene}, Thomas and {Albert}, Lo{\"\i}c and {Nielsen}, Louise D.},
        title = "{PandExo: A Community Tool for Transiting Exoplanet Science with JWST \& HST}",
      journal = {\pasp},
         year = 2017,
        month = jun,
       volume = {129},
       number = {976},
        pages = {064501},
archivePrefix = {arXiv},
 primaryClass = {astro-ph.IM},
       adsurl = {https://ui.adsabs.harvard.edu/abs/2017PASP..129f4501B}
}

@ARTICLE{NIRISS2023,
       author = {{Doyon}, Ren{\'e} and {Willott}, Chris J. and {Hutchings}, John B. and {Sivaramakrishnan}, Anand and {Albert}, Lo{\"\i}c and {Lafreni{\`e}re}, David and {Rowlands}, Neil and {Bego{\~n}a Vila}, M. and {Martel}, Andr{\'e} R. and {LaMassa}, Stephanie and {Aldridge}, David and {Artigau}, {\'E}tienne and {Cameron}, Peter and {Chayer}, Pierre and {Cook}, Neil J. and {Cooper}, Rachel A. and {Darveau-Bernier}, Antoine and {Dupuis}, Jean and {Earnshaw}, Colin and {Espinoza}, N{\'e}stor and {Filippazzo}, Joseph C. and {Fullerton}, Alexander W. and {Gaudreau}, Daniel and {Gawlik}, Roman and {Goudfrooij}, Paul and {Haley}, Craig and {Kammerer}, Jens and {Kendall}, David and {Lambros}, Scott D. and {Ignat}, Luminita Ilinca and {Maszkiewicz}, Michael and {McColgan}, Ashley and {Morishita}, Takahiro and {Ouellette}, Nathalie N.-Q. and {Pacifici}, Camilla and {Philippi}, Natasha and {Radica}, Michael and {Ravindranath}, Swara and {Rowe}, Jason and {Roy}, Arpita and {Roy}, Niladri and {Saad}, Karl and {Sohn}, Sangmo Tony and {Talens}, Geert Jan and {Touahri}, Driss and {Thatte}, Deepashri and {Taylor}, Joanna M. and {Vandal}, Thomas and {Volk}, Kevin and {Wander}, Michel and {Warner}, Gerald and {Zheng}, Sheng-Hai and {Zhou}, Julia and {Abraham}, Roberto and {Beaulieu}, Mathilde and {Benneke}, Bj{\"o}rn and {Ferrarese}, Laura and {Jayawardhana}, Ray and {Johnstone}, Doug and {Kaltenegger}, Lisa and {Meyer}, Michael R. and {Pipher}, Judy L. and {Rameau}, Julien and {Rieke}, Marcia and {Salhi}, Salma and {Sawicki}, Marcin},
        title = "{The Near Infrared Imager and Slitless Spectrograph for the James Webb Space Telescope. I. Instrument Overview and In-flight Performance}",
      journal = {\pasp},
         year = 2023,
        month = sep,
       volume = {135},
       number = {1051},
          eid = {098001},
        pages = {098001},
archivePrefix = {arXiv},
 primaryClass = {astro-ph.IM},
       adsurl = {https://ui.adsabs.harvard.edu/abs/2023PASP..135i8001D}
}

@ARTICLE{NIRSpec2022,
       author = {{Jakobsen}, P. and {Ferruit}, P. and {Alves de Oliveira}, C. and {Arribas}, S. and {Bagnasco}, G. and {Barho}, R. and {Beck}, T.~L. and {Birkmann}, S. and {B{\"o}ker}, T. and {Bunker}, A.~J. and {Charlot}, S. and {de Jong}, P. and {de Marchi}, G. and {Ehrenwinkler}, R. and {Falcolini}, M. and {Fels}, R. and {Franx}, M. and {Franz}, D. and {Funke}, M. and {Giardino}, G. and {Gnata}, X. and {Holota}, W. and {Honnen}, K. and {Jensen}, P.~L. and {Jentsch}, M. and {Johnson}, T. and {Jollet}, D. and {Karl}, H. and {Kling}, G. and {K{\"o}hler}, J. and {Kolm}, M.-G. and {Kumari}, N. and {Lander}, M.~E. and {Lemke}, R. and {L{\'o}pez-Caniego}, M. and {L{\"u}tzgendorf}, N. and {Maiolino}, R. and {Manjavacas}, E. and {Marston}, A. and {Maschmann}, M. and {Maurer}, R. and {Messerschmidt}, B. and {Moseley}, S.~H. and {Mosner}, P. and {Mott}, D.~B. and {Muzerolle}, J. and {Pirzkal}, N. and {Pittet}, J.-F. and {Plitzke}, A. and {Posselt}, W. and {Rapp}, B. and {Rauscher}, B.~J. and {Rawle}, T. and {Rix}, H.-W. and {R{\"o}del}, A. and {Rumler}, P. and {Sabbi}, E. and {Salvignol}, J.-C. and {Schmid}, T. and {Sirianni}, M. and {Smith}, C. and {Strada}, P. and {te Plate}, M. and {Valenti}, J. and {Wettemann}, T. and {Wiehe}, T. and {Wiesmayer}, M. and {Willott}, C.~J. and {Wright}, R. and {Zeidler}, P. and {Zincke}, C.},
        title = "{The Near-Infrared Spectrograph (NIRSpec) on the James Webb Space Telescope. I. Overview of the instrument and its capabilities}",
      journal = {\aap},
         year = 2022,
        month = may,
       volume = {661},
          eid = {A80},
        pages = {A80},
archivePrefix = {arXiv},
 primaryClass = {astro-ph.IM},
       adsurl = {https://ui.adsabs.harvard.edu/abs/2022A&A...661A..80J}
}

@ARTICLE{fastchem1,
       author = {{Stock}, Joachim W. and {Kitzmann}, Daniel and {Patzer}, A. Beate C. and {Sedlmayr}, Erwin},
        title = "{FastChem: A computer program for efficient complex chemical equilibrium calculations in the neutral/ionized gas phase with applications to stellar and planetary atmospheres}",
      journal = {\mnras},
         year = 2018,
        month = sep,
       volume = {479},
       number = {1},
        pages = {865-874},
archivePrefix = {arXiv},
 primaryClass = {astro-ph.EP},
       adsurl = {https://ui.adsabs.harvard.edu/abs/2018MNRAS.479..865S}
}

@ARTICLE{fastchem2,
       author = {{Stock}, Joachim W. and {Kitzmann}, Daniel and {Patzer}, A. Beate C.},
        title = "{FASTCHEM 2 : an improved computer program to determine the gas-phase chemical equilibrium composition for arbitrary element distributions}",
      journal = {\mnras},
         year = 2022,
        month = dec,
       volume = {517},
       number = {3},
        pages = {4070-4080},
archivePrefix = {arXiv},
 primaryClass = {astro-ph.EP},
       adsurl = {https://ui.adsabs.harvard.edu/abs/2022MNRAS.517.4070S}
}

@INPROCEEDINGS{Vogt1999,
       author = {{Vogt}, S.~S. and {Allen}, S.~L. and {Bigelow}, B.~C. and {Bresee}, L. and {Brown}, B. and {Cantrall}, T. and {Conrad}, A. and {Couture}, M. and {Delaney}, C. and {Epps}, H.~W. and {Hilyard}, D. and {Hilyard}, D.~F. and {Horn}, E. and {Jern}, N. and {Kanto}, D. and {Keane}, M.~J. and {Kibrick}, R.~I. and {Lewis}, J.~W. and {Osborne}, J. and {Pardeilhan}, G.~H. and {Pfister}, T. and {Ricketts}, T. and {Robinson}, L.~B. and {Stover}, R.~J. and {Tucker}, D. and {Ward}, J. and {Wei}, M.~Z.},
        title = "{HIRES: the high-resolution echelle spectrometer on the Keck 10-m Telescope}",
    booktitle = {Instrumentation in Astronomy VIII},
         year = 1994,
       editor = {{Crawford}, David L. and {Craine}, Eric R.},
       series = {Society of Photo-Optical Instrumentation Engineers (SPIE) Conference Series},
       volume = {2198},
        month = jun,
        pages = {362},
       adsurl = {https://ui.adsabs.harvard.edu/abs/1994SPIE.2198..362V}
}

@ARTICLE{ACTIN2,
       author = {{Gomes da Silva}, J. and {Santos}, N.~C. and {Adibekyan}, V. and {Sousa}, S.~G. and {Campante}, T.~L. and {Figueira}, P. and {Bossini}, D. and {Delgado-Mena}, E. and {Monteiro}, M.~J.~P.~F.~G. and {de Laverny}, P. and {Recio-Blanco}, A. and {Lovis}, C.},
        title = "{Stellar chromospheric activity of 1674 FGK stars from the AMBRE-HARPS sample. I. A catalogue of homogeneous chromospheric activity}",
      journal = {\aap},
         year = 2021,
        month = feb,
       volume = {646},
          eid = {A77},
        pages = {A77},
archivePrefix = {arXiv},
 primaryClass = {astro-ph.SR},
       adsurl = {https://ui.adsabs.harvard.edu/abs/2021A&A...646A..77G}
}

@ARTICLE{ExofastV2,
       author = {{Eastman}, Jason D. and {Rodriguez}, Joseph E. and {Agol}, Eric and {Stassun}, Keivan G. and {Beatty}, Thomas G. and {Vanderburg}, Andrew and {Gaudi}, B. Scott and {Collins}, Karen A. and {Luger}, Rodrigo},
        title = "{EXOFASTv2: A public, generalized, publication-quality exoplanet modeling code}",
      journal = {arXiv e-prints},
         year = 2019,
        month = jul,
          eid = {arXiv:1907.09480},
        pages = {arXiv:1907.09480},
archivePrefix = {arXiv},
 primaryClass = {astro-ph.EP},
       adsurl = {https://ui.adsabs.harvard.edu/abs/2019arXiv190709480E}
}

@ARTICLE{Fulton2017,
       author = {{Fulton}, Benjamin J. and {Petigura}, Erik A. and {Howard}, Andrew W. and {Isaacson}, Howard and {Marcy}, Geoffrey W. and {Cargile}, Phillip A. and {Hebb}, Leslie and {Weiss}, Lauren M. and {Johnson}, John Asher and {Morton}, Timothy D. and {Sinukoff}, Evan and {Crossfield}, Ian J.~M. and {Hirsch}, Lea A.},
        title = "{The California-Kepler Survey. III. A Gap in the Radius Distribution of Small Planets}",
      journal = {\aj},
         year = 2017,
        month = sep,
       volume = {154},
       number = {3},
          eid = {109},
        pages = {109},
archivePrefix = {arXiv},
 primaryClass = {astro-ph.EP},
       adsurl = {https://ui.adsabs.harvard.edu/abs/2017AJ....154..109F}
}
\bibliographystyle{aasjournal.bst}

\appendix

\counterwithin{table}{section}
\counterwithin{figure}{section}

\section{Acknowledgements}

\begin{acknowledgements}

This work was supported by grant https://doi.org/10.69777/377645 from the Fonds de recherche du Québec and by the Center for Research in Astrophysics of Québec (AstroQuébec).\\
AG  acknowledges support from the Trottier Family Foundation through the Trottier Postdoctoral Fellowship at the Institute for Research on Exoplanets (IREx).\\
RD  acknowledge the financial support of the FRQ-NT through the Centre de recherche en astrophysique du Qu\'ebec as well as the support from the Trottier Family Foundation and the Trottier Institute for Research on Exoplanets.\\
RD  acknowledges support from Canada Foundation for Innovation (CFI) program, the Universit\'e de Montr\'eal and Universit\'e Laval, the Canada Economic Development (CED) program and the Ministere of Economy, Innovation and Energy (MEIE).\\
This work was supported by the Swiss National Science Foundation under grant 2000-1-243028\\
SCB   acknowledges the support from Funda\c{c}\~ao para a Ci\^encia e Tecnologia (FCT) in the form of a work contract through the Scientific Employment Incentive program with reference 2023.06687.CEECIND and DOI 10.54499/2023.06687.CEECIND/CP2839/CT0002.\\
We thank the Swiss National Science Foundation (SNSF) and the Geneva University for their continuous support of our planet search
programmes.\\
This work has been in particular carried out in the frame of the National Centre for Competence in Research PlanetS
supported by the SNSF.\\
This publication makes use of The Data \& Analysis Center for Exoplanets (DACE), which is a facility based
at the University of Geneva dedicated to extra-solar planets data visualisation, exchange, and analysis.\\
ODD  acknowledges support from e-CHEOPS (PEA No 4000142255) and from the Funda\c{c}\~ao para a Ci\^encia e a Tecnologia (FCT) through national funds under the research grant UID/04434/2025 (DOI 10.54499/UID/04434/2025).\\
XD  acknowledges the support from the Swiss National Science Foundation under the grant SPECTRE (No 200021\_215200).\\
JIGH, RR \& ASM  acknowledge financial support from the Spanish Ministry of Science, Innovation and Universities (MICIU) projects PID2020-117493GB-I00 and PID2023-149982NB-I00.\\
JL-B  is funded by the Spanish Ministry of Science and Universities (MICIU/AEI/10.13039/501100011033) PID2023-150468NB-I00\\
GM  acknowledges financial contribution from the INAF GO Large Grant 2023 GAPS-2.\\
We acknowledge financial support from the Agencia Estatal de Investigaci\'on of the Ministerio de Ciencia e Innovaci\'on MCIN/AEI/10.13039/501100011033 and the ERDF “A way of making Europe” through projects PID2021-125627OB-C32 and PID2024-158486OB-C32. This work is supported by the ERC Grant (ERC Advanced Grant SPEAR, GA 101200674).
Funded by the European Union. Views and opinions expressed are however those of the author(s) only and do not necessarily reflect those of the European Union or the  European Research Council Executive Agency (ERCEA). Neither the European Union nor the granting authority can be held responsible for them.\\
RR  acknoweledges the support of the CELESTE EU funding.\\
NCS  acknowledge the support from FCT - Funda\c{c}\~ao para a Ci\^encia e a Tecnologia through national funds by these grants: UIDB/04434/2020, UIDP/04434/2020.\\
Co-funded by the European Union (ERC, FIERCE, 101052347). Views and opinions expressed are however those of the author(s) only and do not necessarily reflect those of the European Union or the European Research Council. Neither the European Union nor the granting authority can be held responsible for them.\\
SGS  acknowledges the support from Funda\c{c}\~ao para a Ci\^encia e Tecnologia (FCT) through national funds under the the Scientific Employment Stimulus programme nº 2024.08008.CEECIND and by the research grant UID/04434/2025 (DOI: 10.54499/UID/04434/2025).\\
MRZO acknowledges financial support from projects PID2022-137241NB-C42 and PID2025-169716NB-C32 from the Spanish ``Ministerio de Ciencia, Innovaci\'on y Universidades''.

\end{acknowledgements}

\onecolumn

\section{Figures}

\begin{figure*}[!h]
    \centering
    \begin{subfigure}{0.49\textwidth}
        \centering
        \includegraphics[width=\linewidth]{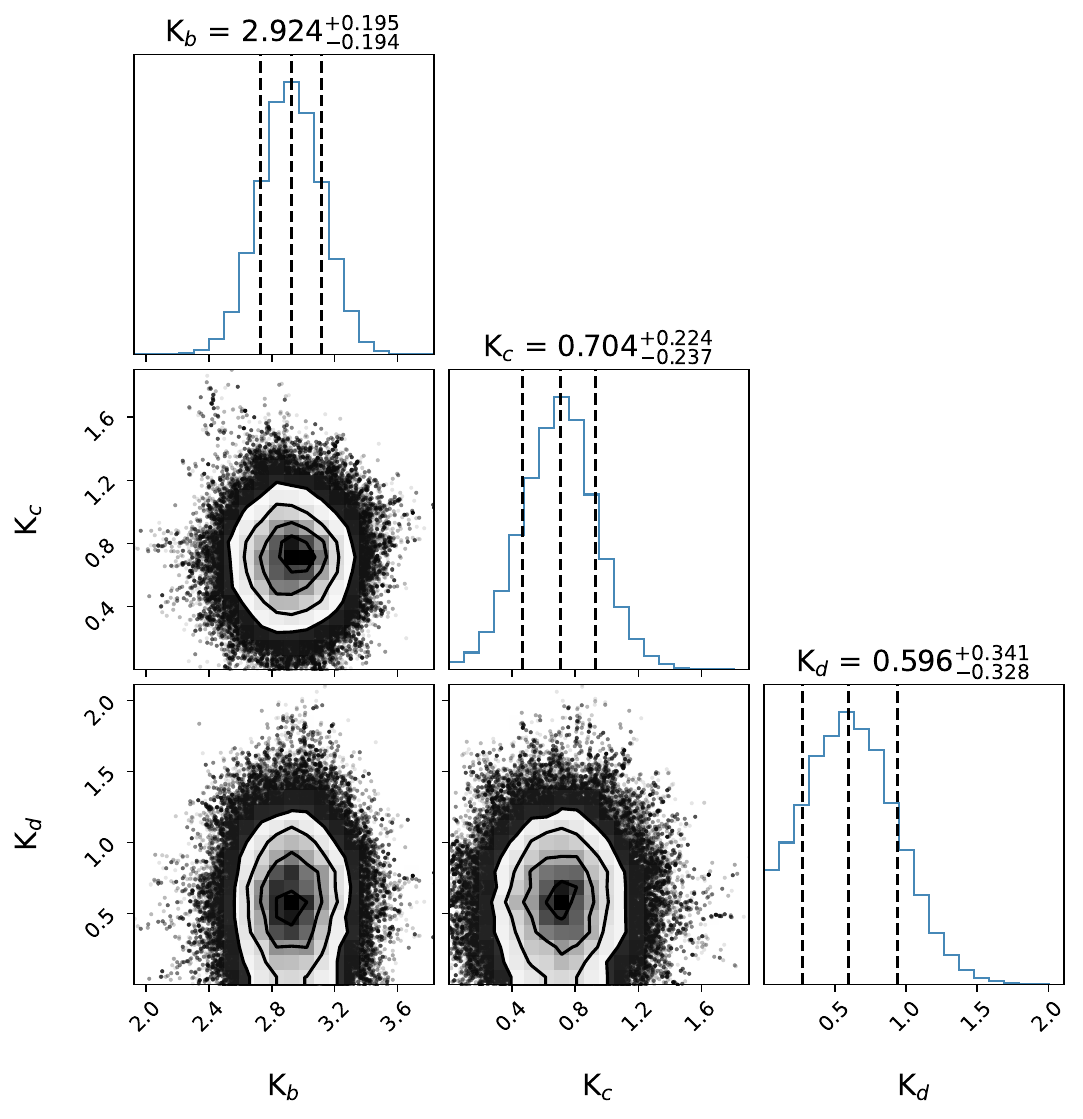}
        \caption{} \label{fig:corner_Kp}
    \end{subfigure}
    \begin{subfigure}{0.49\textwidth}
        \centering
        \includegraphics[width=\linewidth]{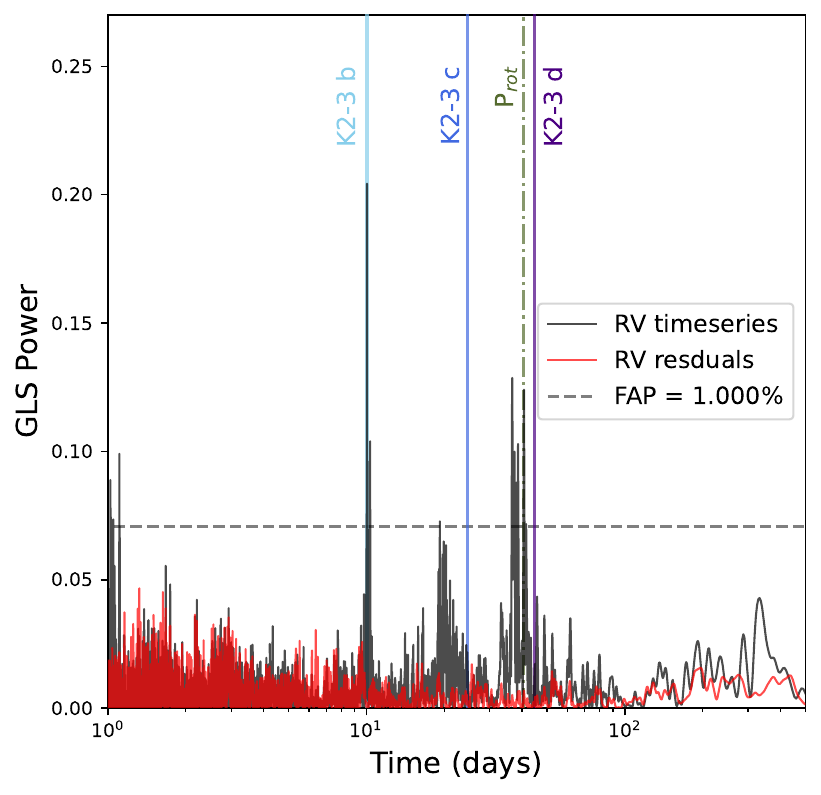}
        \caption{} \label{fig:GLS}
    \end{subfigure}
    \caption{(a) Corner plot for the MGP model featured in Fig.~\ref{fig:full_timeseries_d2v}. While K2-3~b and c are well detected, K2-3~d only has a hint of detection. We note however that it is not an upper limit. An additional 100 ESPRESSO RV measurements are required to constrain the mass of K2-3~d at a 3$\sigma$ level. (b) Generalised Lomb Scargle (GLS,~\citealt{Zechmeister2018-GLS}) periodogram of the same RV timeseries before and after subtraction       of the full (GP + Keplerian) model. Post model subtraction no significant peaks remain. The planet orbital periods and stellar rotation period are marked for reference.}
    \label{fig:corner_GLS}
\end{figure*}

\begin{figure*}[!h]
    \centering
    \begin{subfigure}{0.32\textwidth}
        \centering
        \includegraphics[width=\linewidth]{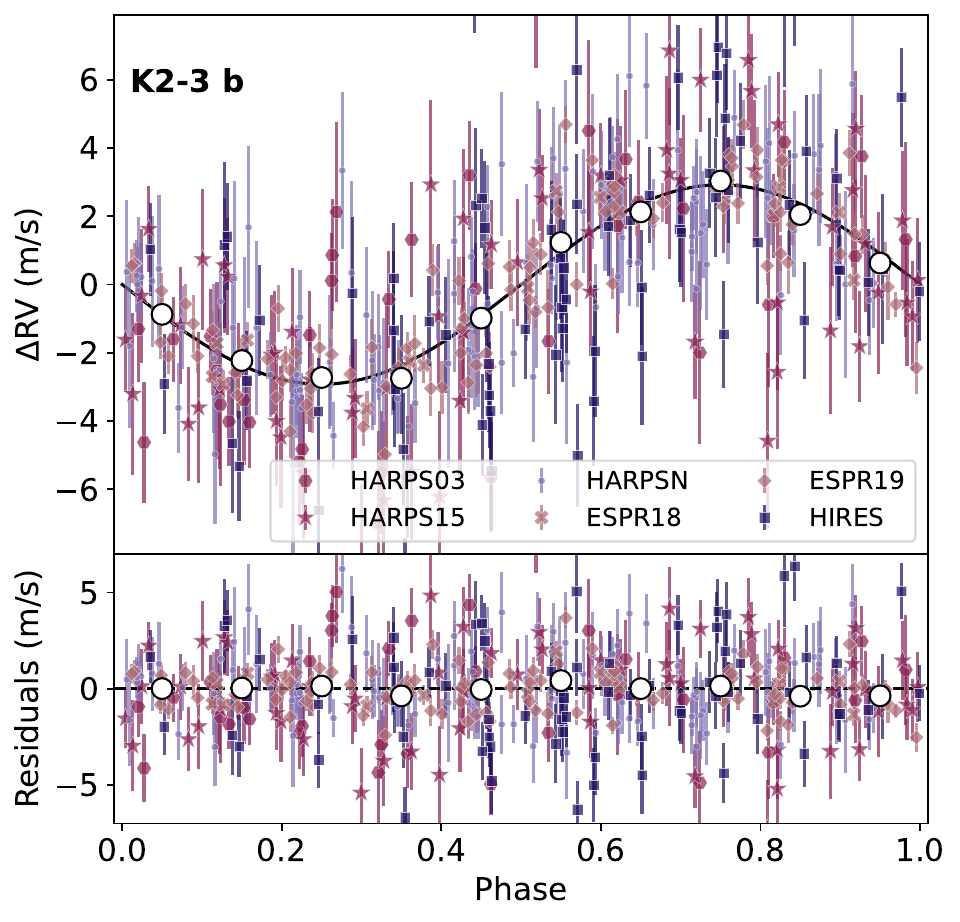}
    \end{subfigure}
    \hfill
    \begin{subfigure}{0.32\textwidth}
        \centering
        \includegraphics[width=\linewidth]{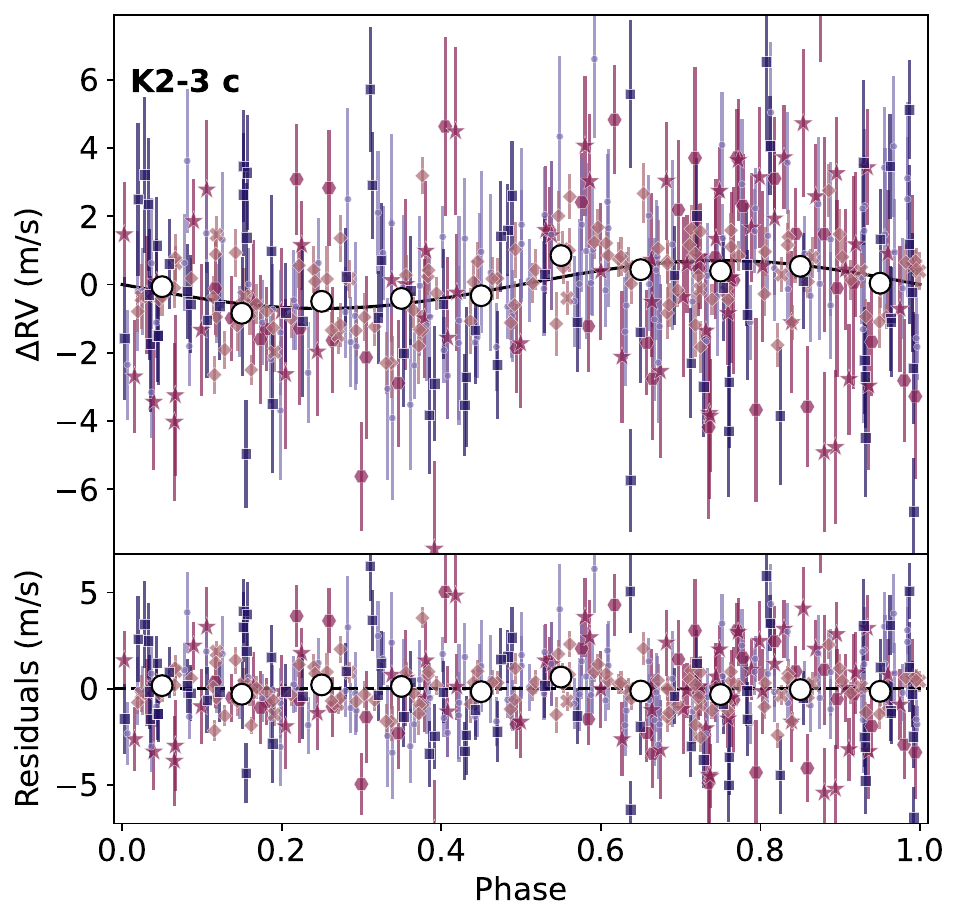}
    \end{subfigure}
    \hfill
    \begin{subfigure}{0.32\textwidth}
        \centering
        \includegraphics[width=\linewidth]{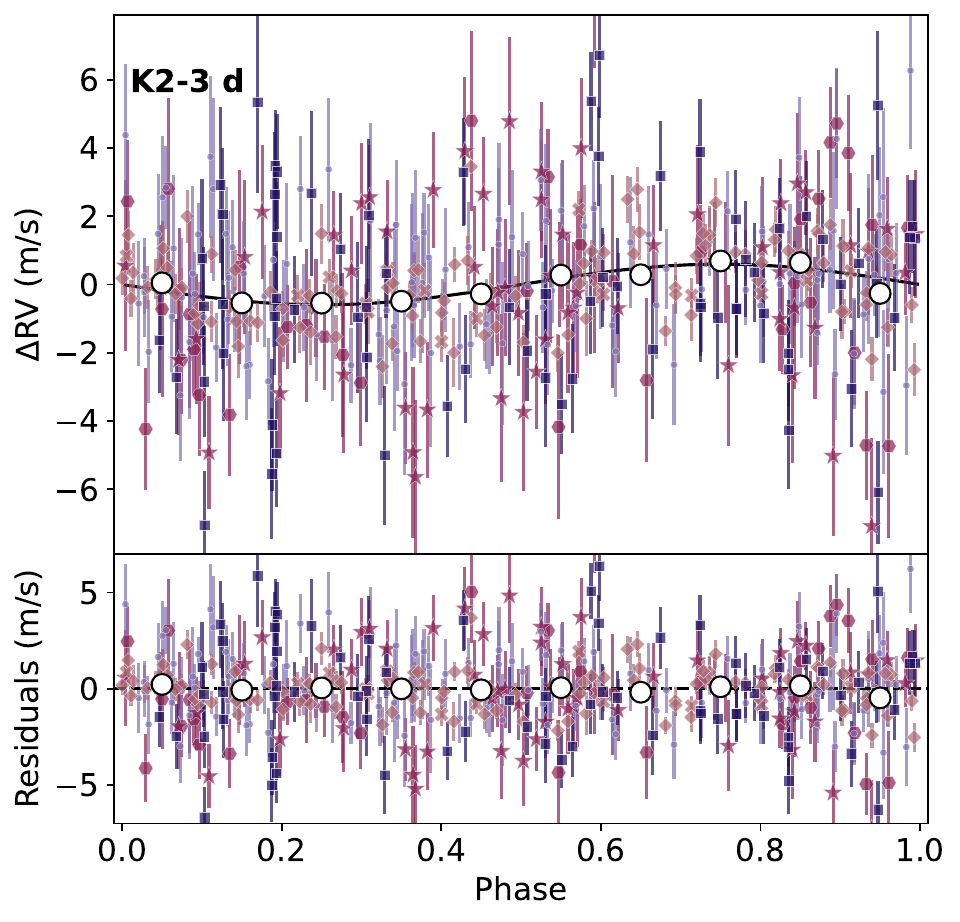}
    \end{subfigure}
    \caption{Phase-folded Keplerian signals of K2-3\,b, c and d. These figures are produced following the model presented in Fig.~\ref{fig:full_timeseries_d2v} using the same colour scheme to represent the different instruments. The binned RVs are presented as the white-face datapoints.}
    \label{fig:rv_phasefold}
\end{figure*}

\begin{figure}[!h]
    \centering
    \includegraphics[width=.6\linewidth]{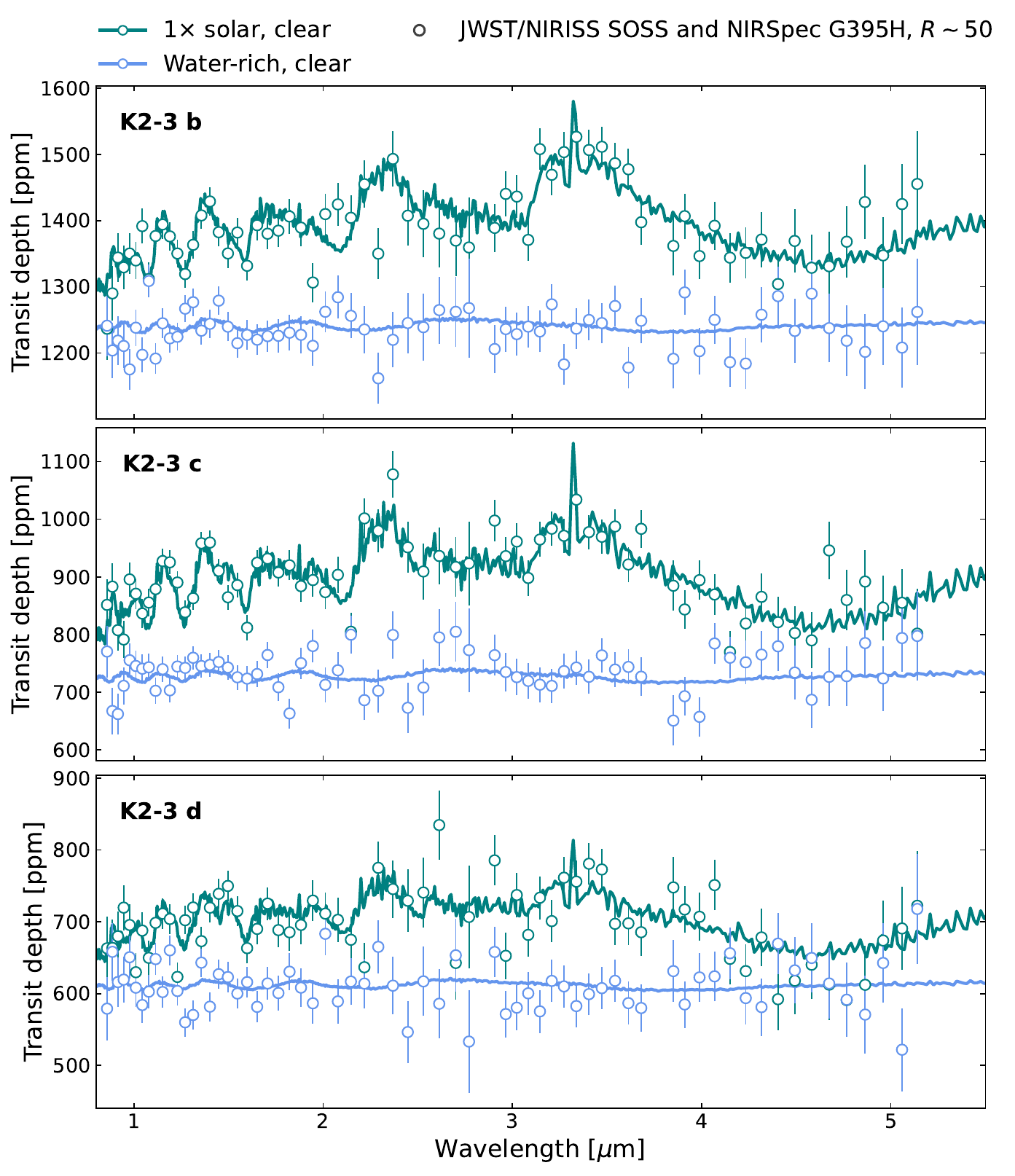}
    \caption{Simulated JWST transmission spectra for K2-3 b, c, and d using the updated system parameters. The solid curves show cloud-free POSEIDON forward models for a solar-composition H$_2$/He atmosphere and an H$_2$O-dominated atmosphere. Open circles and error bars show PandExo simulations of one NIRISS/SOSS transit and one NIRSpec/G395H transit, binned to $R\sim50$.}
    \label{fig:jwst_simu}
\end{figure}

Supplementary figures for this work are shown here. Fig.~\ref{fig:corner_Kp} presents the posterior distribution of the retrieved RV semi-amplitudes from one of our models (Model 1, see Table~\ref{tab:full_model_summary}) and Fig.~\ref{fig:GLS} shows the Generalised Lomb-Scargle (GLS,~\citealt{Zechmeister2018-GLS}) periodogram of the full timeseries before and after subtraction of Model 1. Fig.~\ref{fig:rv_phasefold} has the RV timeseries from Fig.~\ref{fig:full_timeseries_d2v} phasefolded on each planet's orbit. Fig.~\ref{fig:jwst_simu} is an atmospheric simulation of the K2-3 planets featuring two different chemical compositions.

\clearpage
\section{Tables}

Additional tables are presented here. Table~\ref{tab:rv_summary} summarises the dataset used for RV analysis, Table~\ref{tab:planet_summary} presents sampled and derived parameters for all three planets based on the results in this work, and Table~\ref{tab:full_model_summary} contains information regarding all parameters used in all four models of our analysis.

\begin{table}[!h]
    \centering
    \caption{Summary of RV measurements of K2-3.}
    \label{tab:rv_summary}
    \renewcommand{\arraystretch}{1.3}
    \begin{tabularx}{\columnwidth}{@{}X X X X@{}}
        \hline \hline
        \textbf{Instrument} & \textbf{$\bar\sigma_{RV}$ (m/s)} & \textbf{RMS (m/s)}  & \textbf{N$_{meas}$} \\
        \hline
        HARPS03 & 1.78 &  4.51  &  38  \\
        HARPS15 & 1.91 &  4.05   &  64   \\
        HARPS-N & 1.55 &  3.53   &  117   \\
        HIRES$^{\delta}$ & 1.61 &  4.77  &  74   \\
        ESPR18 & 0.54 &  3.03  &  17   \\
        ESPR19 & 0.60 &  4.65  &  99   \\
        \hline
        \multicolumn{4}{@{}p{\columnwidth}@{}}{The median RV uncertainty ($\bar \sigma_{RV}$) calculated after removal of low S/N epochs are presented along with the root mean square (RMS) value for each instrument. The number of epochs (N$_{meas}$) per-instrument used in the analysis is also shown. $^{\delta}$RVs re-computed using LBL for all instruments except for HIRES for which the standard pipeline values were used.}\\
    \end{tabularx}
\end{table}

\begin{table}[!h]
    \centering
    \caption{Parameters of the K2-3 system planets.}
    \label{tab:planet_summary}
    \renewcommand{\arraystretch}{1.3}
    \begin{tabularx}{\columnwidth}{@{}X X X X@{}}
        \hline \hline
        \textbf{Parameter} & \textbf{K2-3\,b} & \textbf{K2-3\,c}  & \textbf{K2-3\,d} \\
        \hline
        $P_\text{orb}$ (days) & $10.0545244\pm0.0000007$ &  $24.6463461\pm0.0000124$  &   $44.5581824\pm0.0001097$  \\
        $t_\text{c}$ (BJD) & $2456813.4200\pm0.0009$ &  $2456812.2778\pm0.0026$  &   $2456826.2245\pm0.0037$  \\
        $K_\text{p}$ (m/s) & $2.92\pm0.18$ &  $0.73\pm0.22$  &  $0.57\pm0.34\,(<1.39)^{\Lambda}$   \\
        $M_\text{p}$ (M$_{\oplus}$) & $6.61\pm0.47$ &  $2.23\pm0.68$   &  $2.12\pm1.27\,(<5.17)^{\Lambda}$   \\
        $R_\text{p}$ (R$_{\oplus}$) & $2.08\pm0.07^{\psi}$ &  $1.58\pm0.05^{\psi}$   &  $1.46\pm0.05^{\psi}$  \\
        $\rho_\text{p}$ (g/cm$^3$) & $4.06\pm0.52$ &  $3.10\pm0.99$  &  $3.77\pm2.29$   \\
        $e$ & $0.08\pm0.05^{\gamma}$ &  0 (fixed)  &  0 (fixed)   \\
        $\omega$ (degrees) & $113\pm17^{\gamma}$ &  0 (fixed)  &  0 (fixed)   \\
        $t_\text{14}$ (hours) & $2.58\pm0.12$ &  $3.58\pm0.06$   &  $4.28\pm0.17$   \\
        $T_\text{eq}$ (K) & $501\pm8$ &  $371\pm6$   &  $305\pm5$   \\
        TSM & $5.76$ &  $5.56$   &  $3.79$   \\
        $v_\text{esc}$ (km/s) & $19.93\pm0.79$ &  $13.26\pm2.02$   &  $13.47\pm4.03$   \\
        \hline
        \multicolumn{4}{@{}p{\columnwidth}@{}}{The parameters presented are: orbital period ($P_\text{orb}$), time of inferior conjugation ($t_\text{c}$), RV semi-amplitude ($K_\text{p}$), planetary mass ($M_\text{p}$), planetary radius ($R_\text{p}$), planetary density ($\rho_\text{p}$), orbital eccentricity ($e$), argument of periastron ($\omega$), transit duration (t$_{14}$), equilibrium temperature ($T_\text{eq}$), transmission spectroscopic metric (TSM,~\citealt{Kempton2018}) and the escape velocity ($v_\text{esc}$). $^{\Lambda}$Due to the low detection significance of K2-3~d, 99th percentile uper limits are also provided. $^{\psi}$These values were taken directly from~\cite{DL2022}. $^{\gamma}$The models with eccentric orbits for K2-3~b are equally statistically favoured as the non-eccentric case.}\\
    \end{tabularx}
\end{table}

\begin{landscape}
\renewcommand{\arraystretch}{1.1}
\setlength{\tabcolsep}{5pt}
\footnotesize
\begin{longtable}[t]{l c c c c c}

\caption{Priors and posteriors for all the parameters fit in our four RV models using Pyaneti. \label{tab:full_model_summary}}\\
\hline\hline
\textbf{Parameter} & \textbf{Prior} & \textbf{Model 1} & \textbf{Model 2} & \textbf{Model 3} & \textbf{Model 4} \\
 & & D2V, $e_b=0$ & D2V, $e_b\neq0$ & H$\alpha$, $e_b=0$ & H$\alpha$, $e_b\neq0$ \\
\hline\\
\endfirsthead

\endhead

\\\hline
\endfoot

\\\hline\hline
\endlastfoot

\textbf{Planet orbital parameters} & & & & & \\
Transit epoch, $t_{0,\text{b}}$ (BJD $-$ 2,400,000)\dotfill & $\mathcal{N}(56813.4202, 0.0009)$ & $56813.4201^{+0.0010}_{-0.0009}$ & $56813.4199^{+0.0009}_{-0.0009}$ & $56813.4203^{+0.0009}_{-0.0009}$ & $56813.4198^{+0.0010}_{-0.0009}$ \\
Orbital period, $P_{\text{b}}$ (days)\dotfill & $\mathcal{N}(10.0545, 0.0001)$ & $10.0545029^{+0.0000004}_{-0.0000005}$ & $10.0545457^{+0.0000009}_{-0.0000009}$ & $10.0545451^{+0.0000004}_{-0.0000004}$ & $10.0545038^{+0.0000009}_{-0.0000008}$ \\
Eccentricity, $e_\text{b}$\dotfill & $\mathcal{U}(0, 0.2)$ & $0.0$ (fixed) & $0.081^{+0.051}_{-0.048}$ & $0.0$ (fixed) & $0.074^{+0.053}_{-0.047}$ \\
Argument of periastron, $\omega_\text{b}$ (radians)\dotfill & $\mathcal{U}(0, 2\pi)$ & $0.0$ (fixed) & $4.114^{+0.746}_{-0.644}$ & $0.0$ (fixed) & $3.791^{+0.583}_{-1.194}$ \\
RV semi-amplitude, $K_\text{b}$ (m/s)\dotfill & $\mathcal{U}(0, 50)$ & $2.924^{+0.196}_{-0.196}$ & $2.924^{+0.192}_{-0.191}$ & $2.930^{+0.176}_{-0.171}$ & $2.934^{+0.190}_{-0.184}$ \\
Transit epoch, $t_{0,\text{c}}$ (BJD $-$ 2,400,000)\dotfill & $\mathcal{N}(56812.2777, 0.0026)$ & $56812.2776^{+0.0026}_{-0.0027}$ & $56812.2779^{+0.0026}_{-0.0026}$ & $56812.2776^{+0.0027}_{-0.0026}$ & $56812.2779^{+0.0026}_{-0.0026}$ \\
Orbital period, $P_\text{c}$ (days)\dotfill & $\mathcal{N}(24.6464, 0.0002)$ & $24.6462192^{+0.0000128}_{-0.0000130}$ & $24.6464712^{+0.0000138}_{-0.0000133}$ & $24.6464723^{+0.0000118}_{-0.0000117}$ & $24.6462207^{+0.0000108}_{-0.0000116}$ \\
Eccentricity, $e_\text{c}$\dotfill & Fixed & $0.0$ & $0.0$ & $0.0$ & $0.0$ \\
Argument of periastron, $\omega_\text{c}$ (radians)\dotfill & Fixed & $0.0$ & $0.0$ & $0.0$ & $0.0$ \\
RV semi-amplitude, $K_\text{c}$ (m/s)\dotfill & $\mathcal{U}(0, 50)$ & $0.704^{+0.226}_{-0.238}$ & $0.677^{+0.238}_{-0.225}$ & $0.729^{+0.196}_{-0.202}$ & $0.726^{+0.200}_{-0.214}$ \\
Transit epoch, $t_{0,\text{d}}$ (BJD $-$ 2,400,000)\dotfill & $\mathcal{N}(56826.2248, 0.0038)$ & $56826.2248^{+0.0038}_{-0.0039}$ & $56826.2236^{+0.0035}_{-0.0036}$ & $56826.2248^{+0.0038}_{-0.0037}$ & $56826.2247^{+0.0036}_{-0.0040}$ \\
Orbital period, $P_\text{d}$ (days)\dotfill & $\mathcal{N}(44.5576, 0.0004)$ & $44.5583986^{+0.0000670}_{-0.0000766}$ & $44.5583954^{+0.0000590}_{-0.0000626}$ & $44.5579582^{+0.0001215}_{-0.0001937}$ & $44.5579859^{+0.0001006}_{-0.0001084}$ \\
Eccentricity, $e_\text{d}$\dotfill & Fixed & $0.0$ & $0.0$ & $0.0$ & $0.0$ \\
Argument of periastron, $\omega_\text{d}$ (radians)\dotfill & Fixed & $0.0$ & $0.0$ & $0.0$ & $0.0$ \\
RV semi-amplitude, $K_\text{d}$ (m/s)\dotfill & $\mathcal{U}(0, 50)$ & $0.596^{+0.343}_{-0.330}$ & $0.664^{+0.332}_{-0.334}$ & $0.475^{+0.410}_{-0.317}$ & $0.561^{+0.397}_{-0.351}$ \\
\\
\textbf{Instrumental parameters} & & & & & \\
Relative systemic RV offset, $\mu_{\text{HARPS03}}$ (m/s)\dotfill & $\mathcal{U}(-109.7, 109.6)$ & $-0.975^{+0.715}_{-0.716}$ & $-0.960^{+0.756}_{-0.755}$ & $-0.826^{+0.744}_{-0.743}$ & $-0.923^{+0.793}_{-0.784}$ \\
Jitter, $\sigma_{w,\text{HARPS03}}$ (m/s)\dotfill & $\mathcal{U}(0, 100)$ & $2.265^{+0.547}_{-0.472}$ & $2.409^{+0.596}_{-0.517}$ & $2.725^{+0.597}_{-0.532}$ & $2.755^{+0.622}_{-0.548}$ \\
Relative systemic RV offset, $\mu_{\text{HARPS15}}$ (m/s)\dotfill & $\mathcal{U}(-110.6, 108.6)$ & $0.610^{+0.615}_{-0.614}$ & $0.542^{+0.610}_{-0.623}$ & $0.572^{+0.545}_{-0.576}$ & $0.528^{+0.564}_{-0.587}$ \\
Jitter, $\sigma_{w,\text{HARPS15}}$ (m/s)\dotfill & $\mathcal{U}(0, 100)$ & $1.717^{+0.433}_{-0.423}$ & $1.815^{+0.427}_{-0.430}$ & $1.745^{+0.415}_{-0.391}$ & $1.758^{+0.416}_{-0.410}$ \\
Relative systemic RV offset, $\mu_{\text{HARPSN}}$ (m/s)\dotfill & $\mathcal{U}(-108.9, 109.1)$ & $-0.595^{+0.504}_{-0.507}$ & $-0.576^{+0.509}_{-0.492}$ & $-0.111^{+0.428}_{-0.445}$ & $-0.198^{+0.470}_{-0.491}$ \\
Jitter, $\sigma_{w,\text{HARPSN}}$ (m/s)\dotfill & $\mathcal{U}(0, 100)$ & $1.090^{+0.294}_{-0.327}$ & $1.079^{+0.301}_{-0.339}$ & $1.193^{+0.297}_{-0.296}$ & $1.194^{+0.294}_{-0.311}$ \\
Relative systemic RV offset, $\mu_{\text{ESPR18}}$ (m/s)\dotfill & $\mathcal{U}(-106.0, 104.9)$ & $1.115^{+1.156}_{-0.997}$ & $1.157^{+1.227}_{-1.161}$ & $1.742^{+0.962}_{-0.830}$ & $1.900^{+1.092}_{-0.916}$ \\
Jitter, $\sigma_{w,\text{ESPR18}}$ (m/s)\dotfill & $\mathcal{U}(0, 100)$ & $1.119^{+0.849}_{-0.596}$ & $1.189^{+1.006}_{-0.636}$ & $1.360^{+0.746}_{-0.539}$ & $1.342^{+0.835}_{-0.584}$ \\
Relative systemic RV offset, $\mu_{\text{ESPR19}}$ (m/s)\dotfill & $\mathcal{U}(-112.6, 111.1)$ & $0.105^{+0.409}_{-0.439}$ & $0.106^{+0.456}_{-0.478}$ & $0.269^{+0.379}_{-0.406}$ & $0.263^{+0.381}_{-0.452}$ \\
Jitter, $\sigma_{w,\text{ESPR19}}$ (m/s)\dotfill & $\mathcal{U}(0, 100)$ & $1.308^{+0.256}_{-0.216}$ & $1.342^{+0.250}_{-0.226}$ & $1.243^{+0.214}_{-0.181}$ & $1.235^{+0.217}_{-0.192}$ \\
Relative systemic RV offset, $\mu_{\text{HIRES}}$ (m/s)\dotfill & $\mathcal{U}(-108.6, 113.7)$ & $-1.065^{+0.670}_{-0.644}$ & $-1.104^{+0.654}_{-0.681}$ & $-1.453^{+0.616}_{-0.637}$ & $-1.511^{+0.647}_{-0.632}$ \\
Jitter, $\sigma_{w,\text{HIRES}}$ (m/s)\dotfill & $\mathcal{U}(0, 100)$ & $3.034^{+0.411}_{-0.385}$ & $3.127^{+0.420}_{-0.389}$ & $2.957^{+0.401}_{-0.350}$ & $2.941^{+0.390}_{-0.359}$ \\
\\
\textbf{MGP hyperparameters} & & & & & \\
GP amplitude, $A_0$ \dotfill & $\mathcal{U}(-100, 100)$ & $-1.398^{+0.301}_{-0.323}$ & $-1.491^{+0.304}_{-0.345}$ & $-0.955^{+0.253}_{-0.304}$ & $-1.033^{+0.272}_{-0.355}$ \\
GP amplitude, $A_1$\dotfill & $\mathcal{U}(-500, 500)$ & $-19.284^{+2.824}_{-3.591}$ & $-19.359^{+2.877}_{-3.575}$ & $-30.592^{+5.540}_{-6.579}$ & $-28.554^{+4.990}_{-6.158}$ \\
GP amplitude, $A_2$\dotfill & $\mathcal{U}(-500, 0)$ & $-7.079^{+0.927}_{-1.118}$ & $-7.392^{+0.977}_{-1.164}$ & $-3.882^{+0.631}_{-0.806}$ & $-3.913^{+0.650}_{-0.798}$ \\
GP amplitude, $A_3$\dotfill & Fixed & $0.0$ & $0.0$ & $0.0$ & $0.0$ \\
GP evolution timescale, $\lambda_e$ (days)\dotfill & $\mathcal{U}(2, 100)$ & $51.437^{+12.854}_{-9.132}$ & $51.762^{+13.635}_{-9.387}$ & $74.215^{+14.500}_{-12.802}$ & $72.226^{+13.868}_{-12.338}$ \\
GP inverse harmonic complexity, $\lambda_p$\dotfill & $\mathcal{U}(0.1, 1.0)$ & $0.594^{+0.077}_{-0.063}$ & $0.601^{+0.081}_{-0.064}$ & $0.775^{+0.108}_{-0.090}$ & $0.737^{+0.094}_{-0.086}$ \\
GP period, $P_{\text{GP}}$ (days)\dotfill & $\mathcal{U}(30, 60)$ & $40.538^{+0.749}_{-0.595}$ & $40.750^{+0.783}_{-0.654}$ & $40.409^{+0.450}_{-0.457}$ & $40.439^{+0.486}_{-0.435}$ \\

\end{longtable}

\begin{tablenotes}
\item\textbf{Notes:} $\mathcal{N}(\mu, \sigma)$ indicates a normal distribution with mean $\mu$ and variance $\sigma^{2}$; $\mathcal{U}(a, b)$ a uniform distribution between $a$ and $b$. HARPS03, HARPS15, HARPSN, ESPR18, ESPR19 and HIRES refer to the spectrographs used for the RV time series. Model 1 and Model 2 use the D2V activity indicator in the MQP model while Model 3 and Model 4 use H$\alpha$ instead.
\end{tablenotes}
\end{landscape}

\end{document}